\documentstyle[12pt,aip]{article}
\newcommand{\be}{\begin{equation}}
\newcommand{\ee}{\end{equation}}
\newcommand{\beq}{\begin{eqnarray}}
\newcommand{\eeq}{\end{eqnarray}}

\newcommand{\k}{\kappa}

\newcommand{\g}[1]{\gamma_{#1}}
\newcommand{\rot}[1]{{\rm \cal R}({#1})}
\newcommand{\tran}[2]{{\rm \cal T}_{({#1},{#2})}}
\newcommand{\gt}[1]{\tilde{\gamma}_{#1}}

\newcommand{\Ft}{\tilde{F}}
\newcommand{\strobe}{{\rm \cal F}}
\newcommand{\F}{\Phi}

\newcommand{\neu}{B_N}
\newcommand{\per}{B_P}
\newcommand{\sg}{\Omega}

\newcommand{\ext}{\tilde{\Omega}}
\newcommand{\rtwo}{{{\rm \bf R}^2}}

\newcommand{\gef}[1]{{\Lambda}_{{#1}}}

\newcommand{\fix}[1]{\mbox{\rm Fix}\,({#1})}
\newcommand{\z}{\zeta}

\newcommand{\eig}[2]{E_{{#1}}({#2})}

\newcommand{\sgd}{D_4}

\newcommand{\tor}{T^2}
\newcommand{\dt}{\tilde{D_4}}
\newcommand{\euc}{{\rm \cal E}(2)}
\newcommand{\otwo}{{\rm  O}(2)}

\newcommand{\norm}[1]{{N}({#1})}
\newcommand{\semi}{\tilde{D_4}\dot{+} T^2}

\newcommand{\rperp}{\vec{r}}

\newtheorem{theorem}{Theorem}[section]
\newtheorem{lemma}{Lemma}[section]

\begin{document}
\baselineskip=14pt    \begin{titlepage}   \vspace*{0.0cm}

\begin{center}
{\bf $\sgd$-symmetric Maps with Hidden Euclidean Symmetry}\end{center}
\begin{center}
John David Crawford \\
Department of Physics and Astronomy\\
University of Pittsburgh\\
Pittsburgh, PA 15260 \\
\end{center}
\vspace{0.5cm}

\centerline{\bf ABSTRACT}
Bifurcation problems in which periodic boundary conditions (PBC) or Neumann
boundary conditions (NBC) are imposed often involve partial differential
equations that have Euclidean symmetry.  In this case posing the bifurcation
problem with either PBC or NBC on a finite domain can lead to a symmetric
bifurcation problem for which the manifest symmetries of the domain do not
completely characterize the constraints due to symmetry on the bifurcation
equations. Additional constraints due to the Euclidean symmetry of the
equations can have a crucial influence on the structure of the bifurcation
equations. An example is the bifurcation of standing waves on the surface of
fluid layer. The Euclidean symmetry of an infinite fluid layer constrains the
bifurcation of surface waves in a finite container with square cross section
because the waves satisfying PBC or NBC can be shown to lie in certain
finite-dimensional fixed point subspaces of the infinite-dimensional problem.
These constraints are studied by analyzing the finite-dimensional vector fields
obtained on these subspaces by restricting the bifurcation equations for the
infinite layer. Particular emphasis is given to determining which bifurcations
might reveal observable effects of the rotational symmetry of the infinite
layer. It turns out that a necessary condition for this possibility to arise is
that the subspace for PBC must carry a reducible representation of the
normalizer subgroup acting on the subspace. This condition can be met in
different ways in both codimension-one and codimension-two bifurcations.
\begin{center}
keywords: bifurcation, symmetry, boundary
conditions, hidden symmetry
\end{center}
\vfill
\begin{center}
Proc. of the Fields Institute Workshop on Pattern Formation and Symmetry
Breaking in PDE's, M. Golubitsky and W. Langford, eds., American Mathematical
Society, to appear.
\end{center}
\begin{center}
August 25, 1993
\end{center}
\end{titlepage}
\tableofcontents
\newpage
\baselineskip=24pt

\section{Introduction}
For  nonlinear maps $\strobe$ on an infinite-dimensional linear space
${E_{\k^2}(\rtwo)}$ that are equivariant with respect to an action of the
Euclidean group $\euc$ on ${E_{\k^2}(\rtwo)}$, we study the finite-dimensional
maps (or vector fields),
\be
\strobe|_{\fix\Sigma},\label{eq:rmap}
\ee
that arise when $\strobe$ is restricted to the fixed point subspaces
$\fix\Sigma$ of certain physically interesting isotropy subgroups $\Sigma$.
These finite-dimensional maps are equivariant with respect to a subgroup of
$\euc$, but in general are also constrained by the original $\euc$ symmetry of
$\strobe$. The restricted map (\ref{eq:rmap}) is thus described as having {\em
hidden} Euclidean symmetry.

Bifurcation problems arising in fluids and other continuum systems motivate
this study; especially recent experimental work on the bifurcation of surface
waves.\cite{sim,cgl} In experiments reported  in 1989, Simonelli and Gollub
observed the appearance of standing waves on the surface of a fluid layer
subjected to a periodic vertical oscillation. This transition is a parametric
instability corresponding to a period doubling bifurcation for the stroboscopic
map.\footnote{The stroboscopic map is the dynamical system defined by recording
the state of the fluid at discrete times separated by one period of the imposed
vibration.} The experiments were performed on a fluid layer in a square
container, but surprisingly the observed standing wave patterns sometimes
failed to break symmetry as predicted from the bifurcation theory based on the
square symmetry of the container. Guided by work on related bifurcation
problems,\cite{gms,fuj,arm1,arm2} Crawford et al. pointed out that under
idealized conditions translation symmetries could constrain the solutions to
the fluid equations describing the waves even though these symmetries did not
respect the boundary conditions of the problem.\cite{craw1} Recent work has
emphasized that the rotational symmetry of the plane also imposes constraints
on the solution in an entirely analogous fashion. \cite{cgl}

The idealized theoretical model describes the motion of an incompressible,
irrotational and inviscid fluid whose state is specified by two fields
$\phi(x,y,z,t)$ and $\eta(x,y,t)$; the velocity potential $\phi$ determines the
velocity of the fluid ${\bf u}=\nabla\phi$ and $\eta$ describes the height of
the free surface $z=\eta(x,y,t)$ at the horizontal position $(x,y)$. Each of
these fields is assumed to satisfy a Neumann boundary condition (NBC), e.g.
$\hat{n}\cdot\nabla\eta(x,y,t)=0$, at the sidewalls of the container.  In
addition, incompressiblity of the fluid implies that $\phi$ satisfies Laplace's
equation, $\nabla^2\phi=0,$
which can be reduced to the two-dimensional Helmholtz equation,
$\nabla_\perp^2\psi(\vec{r})+\k^2\psi(\vec{r})=0$
by separating variables $\phi(x,y,z)=\psi(x,y)\,\cosh\kappa(h+z)$. Here
$\nabla_\perp^2$ is the two-dimensional Laplacian and $h$ denotes the depth of
the layer. A more detailed discussion of this model is given elsewhere;\cite{}
for our purposes it is sufficient to note that the linear wave frequencies
depend on the geometry of the container only through the eigenvalue $\k^2$
which determines the linear degeneracy of the critical eigenvalue near $-1$ for
a given period doubling instability.\cite{cgl}

By virtue of the NBC, solutions to the equations in the physical cross section
$\Omega$ of the container can be smoothly extended by reflection across the
sidewalls to functions on a larger square domain $\ext$. Let $\Omega$
correspond to the domain $0\leq x,y \leq\pi$, then $\ext$ corresponds to
$-\pi\leq x,y \leq\pi$, and the extended functions, defined by
\beq
\eta(-x,y)&\equiv&\eta(x,y)\;\;\;\mbox{\rm for}\;0\leq x\leq\pi
\label{eq:ext1}\\
\eta(x,-y)&\equiv&\eta(x,y)\;\;\;\mbox{\rm for}\;0\leq y\leq\pi,
\eeq
still satisfy the fluid equations and moreover obey periodic boundary
conditions on $\ext$. Hence,  from among {\em all} the solutions to the fluid
equations on $\ext$ with periodic boundary conditions (PBC), the physically
relevant solutions can be picked out by requiring that $\eta$ and $\phi$ are
even under reflection across $x=0$ and $y=0$. In a similar fashion, a solution
on $\ext$ with PBC can always be extended to functions on the entire plane
$\rtwo$ by periodic replication. These functions then satisfy the fluid
equations on all of $\rtwo$. Conversely, if we start with the solutions to the
fluid equations on $\rtwo$, then the initial periodic solutions on $\ext$ can
be picked out simply by requiring $\eta$ (and $\phi$) to satisfy
\be
\eta(x,y)=\eta(x+2\pi,y)=\eta(x,y+2\pi).\label{eq:periodic}
\ee

The second extension to an infinite fluid layer on $\rtwo$ leads to a
bifurcation problem with $\euc$ symmetry. The solutions on $\Omega$ to the
bifurcation problem of direct physical interest are a small subset of the
solutions on $\rtwo$ but clearly must arise in a manner consistent with the
Euclidean symmetry of the extended system. A normal form for the period
doubling bifurcations generating the surface waves in $\Omega$ cannot be
expected to correctly describe these bifurcations unless it incorporates any
constraints due to the ``hidden'' translation and rotation symmetry of the
infinite fluid layer. Such normal forms can be constructed by beginning with
Euclidean symmetric maps appropriate to the infinite layer and then restricting
these maps to the fixed point subspaces associated with the isotropy subgroups
determined by PBC and NBC in (\ref{eq:periodic}) and (\ref{eq:ext1}),
respectively. Some work in this direction has been done to understand the
consequences of the hidden translation symmetry for the normal form of the
stroboscopic map.\cite{craw3} Subsequent experiments by Gollub and Lane have
detected the effects of hidden translation symmetries by breaking these
symmetries in a controlled fashion. Their results agree with the predictions of
the normal form analysis.\cite{cgl}  The analysis of how and when the hidden
Euclidean symmetry affects the normal forms for these bifurcations is extended
in this paper; in particular we investigate whether there are cases in which
one could hope to distinguish between the effects of the hidden translations
and the hidden rotations. It turns out that the specific waves studied in the
experiments of Gollub and Lane are not affected by the hidden rotation
symmetries, but there are codimension-one and codimension-two bifurcations
where it may be possible to detect the presence of the hidden rotational
symmetry. This is one of the interesting experimental issues in this topic.

The Euclidean group $\euc\,:\,\rtwo\rightarrow\rtwo$ is generated by reflection
$\g 2\cdot(x,y)\rightarrow(y,x)$, translations $\tran
ab\cdot(x,y)\rightarrow(x+a,y+b)$, and rotations $\rot\phi$
\be
\rot\phi\cdot(x,y)\rightarrow(x',y')\label{eq:rotdef}
\ee
where
\be
\left(\begin{array}{c}x'\\ y'\end{array}\right) =
\left(\begin{array}{cc}\cos\phi &-\sin\phi\\ \sin\phi & \cos\phi
\end{array}\right)\left(\begin{array}{c}x\\ y\end{array}\right).\nonumber
\ee
The transformation $\gamma\in\euc$ acts on the fluid state $(\phi,\eta)$ in the
usual way
\be
((\gamma\cdot\phi)(\vec{r},z),\;(\gamma\cdot\eta)(\vec{r},z))=
(\phi(\gamma^{-1}\cdot\vec{r},z),\;\eta(\gamma^{-1}\cdot\vec{r},z)).
\label{eq:eucact}
\ee
In terms of this action, the solutions in (\ref{eq:periodic}) satisfying PBC
are those fixed by the subgroup\footnote{The subgroup is specified by listing
its generators.}
$\per = \{\tran {\pm 2\pi}0, \tran 0{\pm 2\pi}\}.$
Similarly the restriction to solutions satisfying NBC on $\Omega$ selects those
that are fixed by the larger subgroup
$\neu = \{\gt 1,\g 3,\tran {\pm 2\pi}0, \tran 0{\pm 2\pi}\}$
where
\beq
\gt 1\cdot(x,y)&\equiv& \g 2 \rot{3\pi/2}\cdot(x,y)=(-x,y)\\
\g 3\cdot(x,y)&\equiv&\g 2\cdot\gt 1\cdot\g 2\cdot(x,y)=(x,-y).\label{eq:g3}
\eeq

The stroboscopic map for the infinite fluid layer commutes with the action of
$\euc$ in (\ref{eq:eucact}). In addition, this map has a fixed point
$(\phi,\eta)=(0,0)$ corresponding to a featureless surface, i.e. the fixed
point is $\euc$ invariant. The linearization of the stroboscopic map at $(0,0)$
thus commutes with the action of $\euc$ and the eigenspaces of the
linearization transform under representations of $\euc$ determined by
(\ref{eq:eucact}).

Let $\gef{\vec{k}}$ denote an eigenvector of the linearized map corresponding
to the critical eigenvalue near $-1$; for the ideal fluid
\be
\gef{\vec{k}}(\vec{r})=
\left(\begin{array}{c}\eta(\k)\\\phi(\k)\end{array}\right)
\;e^{i\vec{k}\cdot\vec{r}},\label{eq:pw}
\ee
and $\gamma\in\euc$ acts on $\gef{\vec{k}}$ by
\be
(\gamma\cdot\gef{\vec{k}})(\vec{r})=\gef{\vec{k}}(\gamma^{-1}\cdot\vec{r}).
\label{eq:linact}
\ee
The fluid fields are real-valued requiring the eigenvector to satisfy
$(\gef{\vec{k}})^\ast=\gef{-\vec{k}}$. The Euclidean invariance of the fixed
point implies that the eigenvalues of the linearization depend only on the
magnitude $|\vec{k}|=\k$ of the wavevector; the specific value of $\k$
associated with the critical eigenvalue varies with the frequency of the
external vibration. For fixed $\vec{k}$, there are two linearly independent
eigenvectors corresponding to eigenvalues that in general are
distinct.\cite{craw3} The infinite-dimensional eigenspace $\eig{\k^2}{\rtwo}$,
associated with the critical eigenvalue of the linearized stroboscopic map,
contains $\gef{\vec{k}}$ and its images (\ref{eq:linact}) under $\euc$.

In the actual experiments, the dissipation in the fluid causes a decay of
transient behavior in the wave motion and simplifies the time-asymptotic
dynamics. When we consider the bifurcation problem on the finite domains $\ext$
(with PBC) and $\Omega$ (with NBC) then the critical eigenspaces
$\eig{\k^2}{\ext}$ and $\eig{\k^2}{\ext}$, respectively, are finite-dimensional
and we assume the dissipation permits a reduced description of the bifurcation
on a finite-dimensional center manifold. Restricting the stroboscopic map to a
center manifold gives a finite-dimensional map which can be locally represented
as a map on the critical eigenspace.  Let $\Ft$ and $f$ denote these
finite-dimensional maps for the domains $\ext$ and $\Omega$ respectively
\beq
\Ft:\eig{\k^2}{\ext}&\rightarrow&\eig{\k^2}{\ext}\label{eq:pbcmap}\\
f:\eig{\k^2}{\Omega}&\rightarrow&\eig{\k^2}{\Omega}\label{eq:nbcmap}.
\eeq

The stroboscopic map for the problem posed on $\ext$ is obtained by simply
applying the map for the infinite layer to fluid states that are spatially
periodic, i.e. those states which are invariant under the subgroup $\per$.
Since the infinite layer map has $\euc$ symmetry, the finite-dimensional map
$\Ft$ on the center manifold for PBC will be constrained by this original
$\euc$ symmetry.  These constraints can enter in two ways. First,
$\eig{\k^2}{\ext}$ is invariant under a subgroup of $\euc$ and $\Ft$ will be
equivariant with respect to this subgroup.  In addition, the form of $\Ft$ may
be further constrained by symmetries in $\euc$ that do {\em not} leave
$\eig{\k^2}{\ext}$ invariant. These latter constraints arise simply from the
fact that $\Ft$ is constructed from maps that are $\euc$-equivariant. The
nature of these latter constraints can be studied by considering the form of
$\euc$-equivariant maps $\strobe$ on the infinite-dimensional eigenspace
$\eig{\k^2}{\rtwo}$
which commute with the representation of $\euc$ carried by the eigenspace.

The connections between the three classes of maps $\strobe$, $\Ft$ and $f$
follow from the isotropy subgroups $\per$ and $\neu$ that characterize the
solutions with PBC and NBC respectively. The eigenspaces are nested
$\eig{\k^2}{\rtwo}\supset\eig{\k^2}{\ext}\supset\eig{\k^2}{\Omega},$
and for $\ext$ and $\Omega$ they are fixed point subspaces:
$\eig{\k^2}{\ext}=\fix\per\label{eq:eigext}$ and
$\eig{\k^2}{\Omega}=\fix\neu.$
The requirement that $\Ft$ and $f$ should be consistent with the Euclidean
symmetry of the infinite layer is formalized by demanding that they be
realizable as the restrictions of an $\euc$-equivariant map $\strobe$ on
$\eig{\k^2}{\rtwo}$ to appropriate subspaces of $\eig{\k^2}{\rtwo}$:
\beq
\Ft&=&\strobe|_{\fix\per}\label{eq:pbcres}\\
f&=&\strobe|_{\fix\neu}=\Ft|_{\fix\neu}.\label{eq:eigom}
\eeq

These restricted maps are not $\euc$-equivariant rather they commute with
certain subgroups of $\euc$. For an isotropy subgroup $\Sigma\subset\euc$, the
{\em normalizer} $\norm\Sigma$ is the maximal subgroup of $\euc$ that leaves
$\fix\Sigma$ invariant, e.g. for $\per\subset\euc$
\be
\norm\per=\{\gamma\in\euc |
\gamma:\fix{\per}\rightarrow\fix{\per}\}.\label{eq:normper}
\ee
Each of the restricted maps is equivariant with respect to the normalizer on
the appropriate fixed point subspace: $\Ft$ commutes with $\norm\per$ and $f$
commutes with $\norm\neu$. This hierarchy of domains, eigenspaces and
symmetries is summarized in Table 1. Note that it is the map $f$ which models
the bifurcations in the physical experiment.

\begin{table}
\begin{center}
Table 1.  Hierarchy of Bifurcation Problems
\end{center}
\vspace{7mm}
\begin{tabular}{lcccc}
\underline{Domain}
&\underline{Boundary Conditions}
&\underline{Eigenspace}
&\underline{Map}
&\underline{Map Symmetry}\\
\\
$\rtwo$ & none & ${E_{\k^2}(\rtwo)}$ &$\strobe$& $\euc$\\
\\
$\ext$ & PBC & ${E_{\k^2}(\ext)}=\fix{\per}$ & $\Ft=\strobe|_{\fix\per}$&
$\norm\per$\\
\\
$\Omega$ & NBC &${E_{\k^2}(\Omega)}=\fix{\neu}$ & $f=\strobe|_{\fix\neu}$&
$\norm\neu$\\
\end{tabular}
\par
\vspace{7 mm}
\underline{Notes:}
\par
PBC are periodic boundary conditions and NBC are Neumann boundary conditions.
\protect \vspace*{\fill}
\end{table}
Although our primary interest is the type of maps $\Ft$ and $f$ that can arise
by restriction from $\strobe$,  we first define ${E_{\k^2}(\rtwo)}$ and
consider the $\euc$-symmetric map $\strobe$. Our focus is on the eigenspaces
${E_{\k^2}(\rtwo)}$ characteristic of codimension-one bifurcations for the
infinite fluid layer.

\subsection{Representations of $\euc$ on $\eig{\k^2}{\rtwo}$}

The eigenvector $\gef{\vec{k}}$ satisfies PBC on $\ext$ if $\vec{k}$ has
integer components: $(k_x,k_y)\in {\cal Z}^2$.  When such eigenfunctions are
extended by periodic replication to $\rtwo$ we get functions that are not
square-integrable.
Following Melbourne\cite{mel}, we define an infinite-dimensional eigenspace
$\eig{\k^2}{\rtwo}$ so that it contains the eigenspace $\eig{\k^2}{\ext}$ and
transforms under the representation determined by (\ref{eq:linact}):
\beq
(\tran ab\cdot \gef{\vec{k}})(\vec{r})&=& e^{-i(ak_x+bk_y)}
\gef{\vec{k}}(\vec{r}),\label{eq:pwtrans}\\
(\rot\phi\cdot\gef{\vec{k}})(\vec{r})&=&\gef{\vec{k}'}(\vec{r})
\label{eq:pwrot}\\
&\mbox{\rm where}&\; \vec{k}'=\rot\phi\cdot\vec{k}\;\;\; \mbox{\rm from}\;
(\ref{eq:rotdef}), \nonumber\\
(\gamma_2\cdot\gef{\vec{k}})(\vec{r})&=&\gef{\vec{k}'}(\vec{r})
\label{eq:pwref}\\
&\mbox{\rm where}& \; \vec{k}'=(k_y,k_x).\nonumber
\eeq

Since Euclidean transformations do not change the length of $|\vec{k}|$, sums
of the form
\be
\Lambda(\vec{r})=\sum_{\vec{k}\in A(\k)}
a(\vec{k})\,\gef{\vec{k}}(\vec{r})\label{eq:genef}
\ee
must be elements of $\eig{\k^2}{\rtwo}$. In this sum $A(\k)$ denotes the circle
of $\vec{k}$ vectors with length $\k$,
and we assume $a(\vec{k})=0$ for all but a finite set of points in $A(\k)$; we
also require $a(\vec{k})^\ast=a(-\vec{k})$ so that $\Lambda(\vec{r})$ is real.
These finite sums form a linear vector space $X(\k)$ which can be given the
norm
\be
\parallel \Lambda(\vec{r})\parallel\equiv \sum_{\vec{k}\in A(\k)} |a(\vec{k})|.
\ee
We define the critical eigenspace $\eig{\k^2}{\rtwo}$ as the closure of $X(\k)$
with respect to this norm
$\eig{\k^2}{\rtwo}\equiv\overline{X(\k)}.$
Since eigenvalues for different values of $\k$ are typically unequal, this
choice for $\eig{\k^2}{\ext}$ is satisfactory for codimension-one bifurcations.
When two (or more) parameters are varied, then mode interactions can occur for
which the sum in (\ref{eq:genef}) should include eigenvectors $\gef{\vec{k}}$
with unequal values of $|\vec{k}|$.
The action of $\euc$ in (\ref{eq:pwtrans}) - (\ref{eq:pwref}) defines a
representation on $\eig{\k^2}{\rtwo}$ in the obvious way. Melbourne has shown
that this representation is absolutely irreducible.\cite{mel}

\begin{theorem} {\rm (Melbourne)} Let $L:\eig{\k^2}{\rtwo}
\rightarrow\eig{\k^2}{\rtwo}$ be a linear map which commutes with the
representation of $\euc$ in {\rm (\ref{eq:pwtrans}) - (\ref{eq:pwref})}, then
$L$ must be a scalar multiple of the identity.
\end{theorem}
\noindent {\em {\bf Proof}.}
\begin{quote}Let
\be
L\gef{\vec{k}}(\vec{r})=\sum_{\vec{k'}\in A(\k)}
c_{\vec{k}}(\vec{k'})\,\gef{\vec{k'}}(\vec{r})\label{eq:linop}
\ee
describe the action of $L$ on $\gef{\vec{k}}(\vec{r})$. When applied to
$\gef{\vec{k}}(\vec{r})$ the assumption that $L$ commutes with an arbitrary
translation ${\rm \cal T}_{\vec{p}}$ requires
$e^{-i\vec{k}\cdot\vec{p}}\, c_{\vec{k}}(\vec{k'})=
e^{-i\vec{k'}\cdot\vec{p}}\,c_{\vec{k}}(\vec{k'})$
for arbitrary $\vec{p}$. Thus $c_{\vec{k}}(\vec{k'})$ must vanish unless
$\vec{k}=\vec{k'}$:
$c_{\vec{k}}(\vec{k'})\equiv\delta_{\vec{k},\vec{k'}}\,C(\vec{k}).$
When $L$ is applied to a general element of $\eig{\k^2}{\rtwo}$, this
simplification implies
\be
L\,\Lambda(\vec{r})=\sum_{\vec{k'}\in A(\k)}
a(\vec{k'})\,C(\vec{k'})\,\gef{\vec{k'}}
\ee
where $\Lambda(\vec{r})$ is given by (\ref{eq:genef}). Acting alone, the
remaining generators $\rot\phi$ and $\g 2$ generate the subgroup $\otwo$, and
the assumption that $\gamma\,L\,\Lambda(\vec{r})=L\,\gamma\,\Lambda(\vec{r})$
for $\gamma\in\otwo$ implies $C(\vec{k'})$ must be an $\otwo$ invariant
function
$C(\gamma\cdot\vec{k'})=C(\vec{k'}).$
The $\otwo$ invariance of $C$ means that it depends only on the magnitude of
$\vec{k'}$ and hence is a $\k$-dependent constant
$C(\vec{k'})=\sigma(\k).$
Thus our linear operator is simply $L=\sigma(\k)\,I$.
 {\bf $\Box$}\end{quote}

\subsection{$\euc$-equivariant maps on $\eig{\k^2}{\rtwo}$}

A second result due to Melbourne that characterizes the nonlinear
$\euc$-equivariant maps on $\eig{\k^2}{\rtwo}$. Let
$\strobe:\eig{\k^2}{\rtwo}\rightarrow\eig{\k^2}{\rtwo}$
be a Euclidean-symmetric homogeneous polynomial map of degree p.
That is we assume
\be
\strobe (\alpha\Lambda)=\alpha^p\strobe (\Lambda)\,\;\;\;{\rm for}\;{\rm
any}\;\alpha\in {\rm C} \label{eq:homo}
\ee
and
\be
\gamma\cdot \strobe (\Lambda)=\strobe (\gamma\cdot \Lambda) \,\;\;\;{\rm
for}\;{\rm any}\;\gamma\in\euc.\label{eq:eucN}
\ee
If we represent $\Lambda$ as in (\ref{eq:genef}), then the homogeneity of
$\strobe $ (\ref{eq:homo}) implies that it must have the form
\be
\strobe (\Lambda)=\sum_{\vec{k}\in A(\k)}\,\gef{\vec{k}}(\vec{r})
\left[\sum_{\vec{k'_1}\in A(\k)}\sum_{\vec{k'_2}\in A(\k)}
\ldots\sum_{\vec{k'_p}\in A(\k)} a(\vec{k'_1})\,a(\vec{k'_2})\,\ldots
a(\vec{k'_p})\,P(\vec{k},\vec{k'_1},\vec{k'_2},\ldots,\vec{k'_p})\right].
\label{eq:pnonlin}
\ee

\begin{theorem} {\rm (Melbourne)} $\strobe$ will have Euclidean symmetry {\rm
(\ref{eq:eucN})}  if $P(\vec{k},\vec{k_1},\vec{k_2},\ldots,\vec{k_p})$
satisfies
\be
P(\vec{k},\vec{k_1},\vec{k_2},\ldots,\vec{k_p})=0\;\;\mbox{\rm if}\;\;
\vec{k}\neq\vec{k_1}+\vec{k_2}+\ldots+\vec{k_p};\label{eq:cond1}
\ee
and
\be
P(\gamma\cdot\vec{k},\gamma\cdot\vec{k_1},\gamma\cdot\vec{k_2},
\ldots,\gamma\cdot\vec{k_p})= P(\vec{k},\vec{k_1},\vec{k_2},\ldots,\vec{k_p})
\;\;\mbox{\rm for all}\;\; \gamma\in \otwo\subset\euc.\label{eq:cond2}
\ee
In addition, if $\strobe $ is real-valued then $P$ must be real-valued.
\end{theorem}
\noindent {\em {\bf Proof}.}
\begin{quote} The Eucliean group $\euc=\otwo\dot{+}T(2)$ is the semi-direct
product of $\otwo$ and the group of translations $T(2)$, and the conditions
(\ref{eq:cond2}) and (\ref{eq:cond1}) correspond to these two components.
\begin{enumerate}
\item The effect of an arbitrary translation ${\rm \cal T}_{\vec{v}}$ on
$\Lambda$ in (\ref{eq:genef}) is to replace $a(\vec{k})$ by
$a(\vec{k})\,e^{-i\vec{k}\cdot\vec{v}}$. Thus ${\rm \cal T}_{\vec{v}}\cdot
\strobe (\Lambda)=\strobe ({\rm \cal T}_{\vec{v}}\cdot\Lambda)$ requires
\be
[e^{-i\vec{v}\cdot(\vec{k_1}+\vec{k_2}+\cdots+\vec{k_p}-\vec{k})}-1]\,P(\vec{k},\vec{k_1},\vec{k_2},\ldots,\vec{k_p})= 0
\ee
for arbitrary $\vec{v}$; hence if
$\vec{k}\neq\vec{k_1}+\vec{k_2}+\ldots+\vec{k_p}$, then
$P(\vec{k},\vec{k_1},\vec{k_2},\ldots,\vec{k_p})=0.$

\item For $\gamma\in\otwo$, we evaluate $\gamma\cdot \strobe $ using the
invariance of the inner product
$\vec{k}\cdot\vec{r}=(\gamma\vec{k})\cdot(\gamma\vec{r})$
\be
\gamma\cdot \strobe (\Lambda)= \sum_{\vec{k}\in A(\k)}\,\gef{\vec{k}}(\vec{r})
\left[\sum_{\vec{k'_1}\in A(\k)} \ldots\sum_{\vec{k'_p}\in A(\k)}
a(\vec{k'_1})\,\ldots
a(\vec{k'_p})\,P(\gamma^{-1}\cdot\vec{k},\vec{k'_1},\ldots,\vec{k'_p}) \right],
\ee
and then from
\be
(\gamma\cdot\Lambda)(\rperp)= \sum_{\vec{k}\in A(\k)}
a(\gamma^{-1}\cdot\vec{k})\,\gef{\vec{k}}(\vec{r}),
\ee
we obtain
\be
\strobe (\gamma\cdot\Lambda)=\sum_{\vec{k}\in A(\k)}\,\gef{\vec{k}}(\vec{r})
\left[\sum_{\vec{k'_1}\in A(\k)} \ldots\sum_{\vec{k'_p}\in A(\k)}
a(\vec{k'_1})\,\ldots
a(\vec{k'_p})\,P(\vec{k},\gamma\cdot\vec{k'_1},\ldots,\gamma\cdot\vec{k'_p})
\right].
\ee
Thus $\gamma\cdot \strobe (\Lambda)=\strobe (\gamma\cdot\Lambda)$ requires that
$P$ is an $\otwo$-invariant function:
$P(\gamma\cdot\vec{k},\gamma\cdot\vec{k_1},\gamma\cdot\vec{k_2},
\ldots,\gamma\cdot\vec{k_p})= P(\vec{k},\vec{k_1},\vec{k_2},\ldots,\vec{k_p}).$

\item Note that for the reflection $\vec{r}\rightarrow-\vec{r}$, this gives
\be
P(-\vec{k},-\vec{k_1},-\vec{k_2}, \ldots,-\vec{k_p})=
P(\vec{k},\vec{k_1},\vec{k_2},\ldots,\vec{k_p});\label{eq:reflct}
\ee
so given a real-valued map $\strobe ^\ast=\strobe $ for which $P$ must satisfy
$P(-\vec{k},-\vec{k_1},-\vec{k_2}, \ldots,-\vec{k_p})^\ast=
P(\vec{k},\vec{k_1},\vec{k_2},\ldots,\vec{k_p})$
the reflection symmetry (\ref{eq:reflct}) implies $P$ is also real-valued.
{\bf $\Box$}
\end{enumerate}
\end{quote}

\section{Periodic boundary conditions and hidden rotation symmetry}

If $\strobe$ is the Euclidean symmetric stroboscopic map on
${E_{\k^2}(\rtwo)}$, then the restriction in (\ref{eq:pbcres}) to $\fix\per$,
yields the stroboscopic map $\Ft$ describing the bifurcation of solutions
satisfying PBC on $\ext$. Since the intersection of the circle $A(\k)$ with the
integer lattice ${\cal Z}^2$,
\be
\tilde{A}(\k)=A(\k)\;\cap\;{\cal Z}^2,\label{eq:akint}
\ee
contains a finite number of points, the eigenspace $E_{\k^2}(\ext)$ will always
be finite-dimensional. Although by construction the restricted map is
equivariant with respect to $\norm\per$, the representation of $\norm\per$ is
not faithful because of the kernel $\per$, and it is the quotient
$\norm\per/\per$ that describes the nontrivial action of $\norm\per$. One can
show quite generally that $\norm\per/\per$ is the semi-direct product $\semi$
of $\dt$ and $\tor$ where $\tor$ denotes the translations
$\tran{\theta_1}{\theta_2}$ with $({\theta_1},{\theta_2})$ taken to be
$2\pi$-periodic variables and $\dt=\{\gt{1},\g{2}\}$ is generated by diagonal
reflection and reflection in $x$.
Depending on $\k^2$, the representation of $\semi$ obtained on $\fix\per$ may
be reducible or irreducible.\cite{cgl} Thus $\Ft$ is always a $\semi$-symmetric
map although the representation involved varies.

It is important to understand whether the hidden Euclidean symmetry of
$\strobe$ imposes constraints on the form of $\Ft$ beyond the obvious
constraint of $\semi$-equivariance. Recall that when a (compact) Lie group
$\Gamma$ acts on a linear vector space $V$, the set of all (smooth)
$\Gamma$-symmetric vector fields on $V$ has the structure of a module over the
ring of $\Gamma$-invariant functions on $V$. Let $R(\Gamma)$ and
$\vec{M}(\Gamma)$ denote this ring and module, respectively.\cite{gss} For
$\Gamma=\semi$ acting on $V=E_{\k^2}(\ext)$, the set of all $\semi$-equivariant
maps is the module $\vec{M}(\semi)$, and the restriction in (\ref{eq:pbcres})
always yields an element of $\vec{M}(\semi)$. The issue of hidden constraints
poses the inverse problem: if $\Ft\in\vec{M}(\semi)$ is taken to be an
arbitrary $\semi$-equivariant map on $E_{\k^2}(\ext)$ can one always obtain
such an $\Ft$ by the restriction in (\ref{eq:pbcres})? The answer depends on
whether or not the representation of $\semi$ carried by ${E_{\k^2}(\ext)}$ is
irreducible.

The (non-trivial) irreducible representations of $\semi$ have dimension four or
eight and may be classified by two mode numbers $\vec{k}=(l,n)$ with $l\geq
n\geq 0$ giving the components of a specific wavevector in $\tilde{A}(\k)$
.\cite{dg} There are two sets of distinct four-dimensional irreducible
representations:
\begin{enumerate}
\item For $l=n>0$, $\tilde{A}(\k)=\{\pm\vec{k_1},\pm\vec{k_2}\}$ where
$\vec{k_1}=(l,l)$ and $\vec{k_2}=(l,-l)$.  Then $\k^2=2l^2$ and
\be
{E_{\k^2}(\ext)}= \{z_1\gef{\vec{k_1}}(\vec{r}) + z_2\gef{\vec{k_1}}(\vec{r}) +
cc)|\;(z_1,z_2)\in {\bf C}^2\},\label{eq:asubtll}
\ee
carries the representation of $\semi$ generated by
\beq
\gt{1}\cdot(z_1,z_2)&=&(\overline{z}_2,\overline{z}_1)\label{eq:4drepa}\\
\g{2}\cdot(z_1,z_2)&=&(z_1,\overline{z}_2)\label{eq:4drepb}\\
\tran{\theta_1}{\theta_2}\cdot(z_1,z_2)&=&
(e^{-il(\theta_1+\theta_2)}z_1,e^{-il(\theta_1-\theta_2)}z_2).
\label{eq:4drepc}
\eeq

\item For $l>n=0$, $\tilde{A}(\k)=\{\pm\vec{k_1},\pm\vec{k_2}\}$ where
$\vec{k_1}=(l,0)$ and $\vec{k_2}=(0,l)$. Then $\k^2=l^2$ and
\be
{E_{\k^2}(\ext)}= \{z_1\gef{\vec{k_1}}(\vec{r}) + z_2\gef{\vec{k_2}}(\vec{r}) +
cc)|\;(z_1,z_2)\in {\bf C}^2\},\label{eq:asubtl0}
\ee
carries the representation of $\semi$ generated by
\beq
\gt{1}\cdot(z_1,z_2)&=&(\overline{z}_1,{z}_2)\label{eq:l0repa}\\
\g{2}\cdot(z_1,z_2)&=&(z_2,{z}_1)\label{eq:l0repb}\\
\tran{\theta_1}{\theta_2}\cdot(z_1,z_2)&=&
(e^{-il\theta_1}z_1,e^{-il\theta_2}z_2). \label{eq:l0repc}
\eeq
\end{enumerate}
Note that $\tran{\pi/l}{\pi/l}$ is in the kernel of (\ref{eq:asubtll})  but not
of (\ref{eq:asubtl0}), hence the two sets of representations must be
inequivalent.

The eight-dimensional representations correspond to $l>n>0$ with
$\tilde{A}(\k)=\{\pm\vec{k_1},\pm\vec{k_2},\pm\vec{k_3},\pm\vec{k_4}\}$ where
$\vec{k_1}=(l,n)$, $\vec{k_2}=(l,-n)$, $\vec{k_3}=(n,l)$, and
$\vec{k_4}=(n,-l)$. Now $\k^2=l^2+n^2$ and
\be
{E_{\k^2}(\ext)} = \{(z_1\gef{\vec{k_1}}(\vec{r}) + z_2\gef{\vec{k_2}}(\vec{r})
+ z_3\gef{\vec{k_3}}(\vec{r}) + z_4\gef{\vec{k_4}}(\vec{r}) +
cc)|\;(z_1,z_2,z_3,z_4)\in {\bf C}^4\}\label{eq:subt}
\ee
transforms under the following representation of $\semi$:
\beq
\gt{1}\cdot(z_1,z_2,z_3,z_4)&=&
(\overline{z}_2,\overline{z}_1,\overline{z}_4,\overline{z}_3)\label{eq:8drepa}\\
\g{2}\cdot(z_1,z_2,z_3,z_4)&=&
(z_3,\overline{z}_4,z_1,\overline{z}_2)\label{eq:8drepb}\\
\tran{\theta_1}{\theta_2}\cdot(z_1,z_2,z_3,z_4)&=&
(e^{-i(l\theta_1+n\theta_2)}z_1,e^{-i(l\theta_1-n\theta_2)}z_2,
e^{-i(n\theta_1+l\theta_2)}z_3,e^{-i(n\theta_1-l\theta_2)}z_4).\label{eq:8drepc}
\eeq

We shall assume that $(l,n)$ are relatively prime although this is not required
for irreducibility.  Our analysis will depend only on the invariant functions
$R(\semi)$ and equivariant vector fields $\vec{M}(\semi)$ for a given
representation. For two irreducible representations $(l,n)$ relatively prime
and $(l'=ml,n'=mn)$, the sets $R(\semi)$ and $\vec{M}(\semi)$ are the
same.\cite{gss} Our main result is the following theorem.

\begin{theorem} Assume ${E_{\k^2}(\ext)}$ carries an irreducible representation
of $\semi$, and
$\Ft:{E_{\k^2}(\ext)}\rightarrow{E_{\k^2}(\ext)}$
is a $\semi$-equivariant map, then there is a $\euc$-equivariant map
$\strobe:{E_{\k^2}(\rtwo)}\rightarrow{E_{\k^2}(\rtwo)}$
such that
\be
\Ft = \strobe |_{{E_{\k^2}(\ext)}}.\label{eq:4dpbc}
\ee
\end{theorem}
\noindent {\em {\bf Proof}.}
\begin{quote}  The theorem follows from lemma II.1 and lemma II.7 below which
treat the four-dimensional and eight-dimensional cases respectively.
{\bf $\Box$}\end{quote}
The content of this theorem is not new to experts in the subject but it seems
that a proof has never been published.
\subsection{Four-dimensional irreducible representations of $\semi$}
For the four-dimensional representations, Gomes has given a general
representation of $\semi$-equivariant maps $\Ft$ on
${E_{\k^2}(\ext)}$:\cite{gomes,gomes2}
\be
\Ft(z_1,z_2)= \tilde{p}(\tilde{N},\tilde{\rho}) \left(\begin{array}{c}
z_1\\z_2\end{array}\right) + \tilde{q}(\tilde{N},\tilde{\rho})
\left(\begin{array}{c} z_1|z_2|^2\\z_2|z_1|^2\end{array}\right) +
\tilde{r}(\tilde{N},\tilde{\rho}) \left(\begin{array}{c}
z_1|z_1|^2\\z_2|z_2|^2\end{array}\right)\label{eq:llnf}
\ee
where $\tilde{p}$, $\tilde{q}$, and $\tilde{r}$ are arbitrary
real-valued functions of the basic $\semi$ invariants
$\tilde{N}=|z_1|^2 + |z_2|^2$ and $\tilde{\rho}=|z_1|^2|z_2|^2.$
By simply checking that the three basic equivariants in (\ref{eq:llnf}) and the
basic invariants $\tilde{N}$ and $\tilde{\rho}$ can be extended to Euclidean
equivariants and invariants on ${E_{\k^2}(\rtwo)}$ one can prove the existence
of an $\euc$-equivariant $\strobe$ on ${E_{\k^2}(\rtwo)}$ which restricts to
the general $\Ft$ as in (\ref{eq:pbcres}). However, when we consider the
eight-dimensional irreducible representation and higher-dimensional reducible
representations associated with mode interactions, it becomes increasingly
laborious to enumerate such a Hilbert basis for the $\semi$ invariants and
generators for the $\semi$ equivariants. Thus in this simpler case we
illustrate an alternative approach which requires only a study of the $\tor$
invariants of a given representation of $\semi$.

\begin{lemma} Assume ${E_{\k^2}(\ext)}$ carries a four-dimensional irreducible
representation of $\semi$ as in {\rm (\ref{eq:asubtll})} or {\rm
(\ref{eq:asubtl0})}, and $\Ft:{E_{\k^2}(\ext)}\rightarrow{E_{\k^2}(\ext)}$
is a $\semi$-equivariant map, then there is a $\euc$-equivariant map
$\strobe:{E_{\k^2}(\rtwo)}\rightarrow{E_{\k^2}(\rtwo)}$
such that
$\Ft = \strobe |_{{E_{\k^2}(\ext)}}.$
\end{lemma}
\noindent {\em {\bf Proof}.}
\begin{quote}
\begin{enumerate}
\item The reflections $\gt 1$ and $\g 2$ imply $\Ft$ may be written in terms of
a single function $F_1(z_1,z_2)$. Translation symmetry of $\Ft$ then implies
that $F_1$ is simply related to the $\tor$ invariants. Let
\be
\Ft(z_1,z_2)= \left(\begin{array}{c}
F_1(z_1,z_2)\\F_2(z_1,z_2)\end{array}\right);
\label{eq:4dmap}
\ee
for representations (\ref{eq:asubtll}) associated with equal wavenumbers
$\gt 1\cdot\Ft(z_1,z_2)= \Ft(\gt 1\cdot(z_1,z_2))$ requires
$F_2(z_1,z_2)= F_1(z_2^\ast,z_1^\ast)^\ast$
and
$\g 2\cdot\Ft(z_1,z_2)= \Ft(\g 2\cdot(z_1,z_2))$ implies
\be
\begin{array}{c}
F_1(z_1,z_2)=F_1(z_1,z_2^\ast)\\
F_1(z_1,z_2)=F_1(z_1^\ast,z_2^\ast)^\ast.
\end{array}\label{eq:4dref2}
\ee
Translation symmetry, $\tran{\theta_1}{\theta_2}\cdot\Ft
(z_1,z_2)=\Ft(\tran{\theta_1}{\theta_2}\cdot (z_1,z_2))$, requires
\be
e^{-il(\theta_1+\theta_2)}F_1(z_1,z_2)=
F_1(e^{-il(\theta_1+\theta_2)}z_1,e^{-il(\theta_1-\theta_2)}z_2);
\ee
which states that $z_1^\ast\,F_1(z_1,z_2)$ is a translation-invariant function.
Thus for this representation all equivariant maps have the form
\be
\Ft(z_1,z_2)=\left(\begin{array}{c}
F_1(z_1,z_2)\\F_1(z_2,z_1)\end{array}\right)
\label{eq:4dmapll}
\ee
where $F_1$ satisfies (\ref{eq:4dref2}) and $(z_1^\ast\,F_1)$ is
$\tor$-invariant.
\item For representations (\ref{eq:asubtl0}) associated with zero wavenumbers,
the same considerations show that $\Ft$ must have the form
\be
\Ft(z_1,z_2)=\left(\begin{array}{c}
F_1(z_1,z_2)\\F_1(z_2,z_1)\end{array}\right)
\label{eq:4dmapl0}
\ee
where $F_1$ satisfies
$F_1(z_1,z_2)=F_1(z_1,z_2^\ast)=F_1(z_1^\ast,z_2^\ast)^\ast$
and $(z_1^\ast\,F_1)$ is $\tor$-invariant.

\item For both types of irreducible representations, the quadratic invariants
$\sigma_1=|z_1|^2$ and $\sigma_2=|z_2|^2$
provide a Hilbert basis for $\tor$-invariant functions. Thus, in both cases,
the general monomial $F_1$ defining a homogeneous symmetric map $\Ft(z_1,z_2)$
of degree $p$ is
\be
F_1(z_1,z_2)=z_1\,\sigma_1^m\,\sigma_2^n\label{eq:F1}
\ee
with $p=2(m+n)+1$. (The form of the resulting map $\Ft$ depends on the
representation considered.)

\item Now consider the form of the $\semi$-equivariant map obtained when we
restrict an $\euc$-equivariant map $\strobe(\Lambda)$ on ${E_{\k^2}(\rtwo)}$ to
the four-dimensional eigenspace ${E_{\k^2}(\ext)}$ defined in
(\ref{eq:asubtll}) or (\ref{eq:asubtl0}). In either case the restriction is
obtained in same manner: $\Lambda\in {E_{\k^2}(\rtwo)}$ belongs to
${E_{\k^2}(\ext)}$ if $a(\vec{k})$ in (\ref{eq:genef}) has the form
\be
a(\vec{k})=z_1\delta_{\vec{k},\vec{k_1}}+z_2\delta_{\vec{k},\vec{k_2}}
+z_1^\ast\delta_{\vec{k},\vec{-k_1}}+z_2^\ast\delta_{\vec{k},\vec{-k_2}}
\label{eq:restrict}
\ee
where the definition of $\vec{k_1}$ and $\vec{k_2}$ depends on the
representation. Inserting $a(\vec{k})$ from (\ref{eq:restrict}) into the right
hand side of (\ref{eq:pnonlin}) gives
\be
\strobe|_{{E_{\k^2}(\rtwo)}}=\sum_{\vec{k'}\in A(\k)}\,a'(\vec{k'})
\,\gef{\vec{k'}}(\vec{r})
\ee
where
\be
a'(\vec{k'})=
\sum_{\vec{k'_1}\in \tilde{A}(\k)} \ldots\sum_{\vec{k'_p}\in \tilde{A}(\k)}
a(\vec{k'_1})\,a(\vec{k'_2})\,\ldots
a(\vec{k'_p})\,P(\vec{k'},\vec{k'_1},\vec{k'_2},\ldots,\vec{k'_p})
\label{eq:resnf}
\ee
with $\tilde{A}(\k)=\{\pm\vec{k_1},\pm\vec{k_2}\}$.
Since $a'(\vec{k'})$, defined in this way, must also have the form
(\ref{eq:restrict}),
we can read off the function $F_1(z_1,z_2)$ for the resulting
$\semi$-equivariant map by setting $\vec{k'}=\vec{k_1}$ in (\ref{eq:resnf}):
\be
F_1(z_1,z_2)=\sum_{\vec{k'_1}\in \tilde{A}(\k)}\ldots\sum_{\vec{k'_p}\in
\tilde{A}(\k)} a(\vec{k'_1})\,a(\vec{k'_2})\,\ldots
a(\vec{k'_p})\,P(\vec{k_1},\vec{k'_1},\vec{k'_2}, \ldots,\vec{k'_p}).
\label{eq:F1nf}
\ee

\item Comparing (\ref{eq:F1}) and (\ref{eq:F1nf}) shows that to prove that a
given $\semi$-equivariant map can be obtained  by restriction from a
$\euc$-symmetric map on ${E_{\k^2}(\rtwo)}$, it is sufficient to check the
existence of $P(\vec{k_1},\vec{k'_1},\vec{k'_2},\ldots,\vec{k'_p})$ satisfying
both (\ref{eq:cond1}) and (\ref{eq:cond2}) and yielding
\be
z_1\,\sigma_1^m\,\sigma_2^n=\sum_{\vec{k'_1}\in \tilde{A}(\k)}
\ldots\sum_{\vec{k'_p}\in \tilde{A}(\k)} a(\vec{k'_1})\,\ldots
a(\vec{k'_p})\,P(\vec{k_1},\vec{k'_1},\ldots,\vec{k'_p})\label{eq:cond}
\ee
for arbitrary integers $m\geq0$ and $n\geq0$. The $O(2)$ invariance of $P$
(\ref{eq:cond2}) can be manifestly assured by constructing it from the inner
products of its arguments. The condition for $\tor$-equivariance
(\ref{eq:cond1}) can be checked once $P$ satisfies (\ref{eq:cond}).

\item For both four-dimensional representations, we have
\be
\left(\begin{array}{cc}
\vec{k_1}\cdot\vec{k_1}&\vec{k_1}\cdot\vec{k_2}\\
\vec{k_2}\cdot\vec{k_1}&\vec{k_2}\cdot\vec{k_2}
\end{array}\right)=\left(\begin{array}{cc}
\k^2&0\\
0&\k^2
\end{array}\right)
\ee
so that up to an overall sign there are only two inner products that occur for
vectors in $\tilde{A}(\k)$. Define
$I_1(\vec{k},\vec{q})=\delta_{\vec{k}\cdot \vec{q},\k^2}$ and
$I_2(\vec{k},\vec{q})=\delta_{\vec{k}\cdot \vec{q},0}$
to describe these two products; using $I_1$ and $I_2$, we can easily construct
$P$ to satisfy (\ref{eq:cond}). A few examples illustrate the point. For
$m=n=0$, set
$P(\vec{k_1},\vec{k'_1})=I_1(\vec{k_1},\vec{k'_1})$
which yields
\be
z_1=\sum_{\vec{k'_1}\in A(\k)} a(\vec{k'_1})\,P(\vec{k_1},\vec{k'_1}),
\ee
and clearly satisfies the conditions (\ref{eq:cond1}) and (\ref{eq:cond2})
required by Euclidean symmetry. We add a factor of $\sigma_1$ by choosing
\be
P(\vec{k_1},\vec{k'_1},\vec{k'_2},\vec{k'_3})=I_1(\vec{k_1},\vec{k'_1})\,
I_1(\vec{k_1},\vec{k'_2})\,I_1(\vec{k'_2},\vec{-k'_3})\label{eq:pfcn1}
\ee
which gives
\be
z_1\,\sigma_1^2=\sum_{\vec{k'_1}\in \tilde{A}(\k)}\sum_{\vec{k'_2}\in
\tilde{A}(\k)} \sum_{\vec{k'_3}\in \tilde{A}(\k)}
a(\vec{k'_1})\,a(\vec{k'_2})\,a(\vec{k'_3})
\,P(\vec{k_1},\vec{k'_1},\vec{k'_2},\vec{k'_3}).
\ee
Similarly choosing
\be
P(\vec{k_1},\vec{k'_1},\vec{k'_2},\vec{k'_3})=I_1(\vec{k_1},\vec{k'_1})\,
I_2(\vec{k_1},\vec{k'_2})\,I_1(\vec{k'_2},\vec{-k'_3})/2\label{eq:ex3}
\ee
gives
\be
z_1\,\sigma_2^2=\sum_{\vec{k'_1}\in \tilde{A}(\k)} \sum_{\vec{k'_2}\in
\tilde{A}(\k)}
\sum_{\vec{k'_3}\in \tilde{A}(\k)} a(\vec{k'_1})\,a(\vec{k'_2})\,a(\vec{k'_3})
\,P(\vec{k_1},\vec{k'_1},\vec{k'_2},\vec{k'_3}).
\ee
For fixed $\vec{k_1}$ there are two solutions to $\vec{k_1}\perp\vec{k'_2}$ so
we need the factor of $1/2$ in (\ref{eq:ex3}). The requirement of translation
symmetry (\ref{eq:cond1}) is obviously satisfied for the functions in
(\ref{eq:pfcn1}) and (\ref{eq:ex3}). By appending additional factors of $I_1$
and $I_2$ with appropriately specified arguments one can arrange for an
arbitrary number of factors of $\sigma_1$ and $\sigma_2$, respectively, without
losing the translation symmetry (\ref{eq:cond1}). {\bf $\Box$}
\end{enumerate}
\end{quote}

\subsection{Eight-dimensional irreducible representations of $\semi$}

The corresponding result for the eight-dimensional representations is proved
similarly. The analysis of a $\semi$ equivariant can be reduced to the study of
the $\tor$ invariants for the representation. The argument is longer however
because the ring of $\tor$ invariants is more complicated.

For the representation (\ref{eq:8drepa}) - (\ref{eq:8drepc}) with $l>n>0$
relatively prime, let ${\bf z}\equiv(z_1,z_2,z_3,z_4)$ and consider an
arbitrary $\semi$ equivariant map
\be
\Ft({\bf z})=\left(\begin{array}{c} F_1({\bf z})\\ F_2({\bf z})\\  F_3({\bf
z})\\ F_4({\bf z})\end{array}\right).
\label{eq:8dmap}
\ee
The reflection symmetry, $\gt 1\cdot\Ft ({\bf z})=\Ft(\gt 1\cdot ({\bf z}))$,
requires
$F_2({\bf z})=F_1(\gt 1\cdot{\bf z})^\ast$ and $F_4({\bf z})=F_3(\gt 1\cdot{\bf
z})^\ast$,
and
$\g 2\cdot\Ft ({\bf z})=\Ft(\g 2\cdot{\bf z})$ implies
$F_3({\bf z})=F_1(\g 2\cdot{\bf z})$ and
\be
F_1({\bf z})=F_1({\bf z}^\ast)^\ast.\label{eq:8dref4}
\ee
The translation symmetry, $\tran{\theta_1}{\theta_2}\cdot\Ft ({\bf
z})=\Ft(\tran{\theta_1}{\theta_2}\cdot ({\bf z}))$, requires
$e^{-i(l\theta_1+n\theta_2)}F_1({\bf z}) = F_1(\tran{\theta_1}{\theta_2}\cdot
{\bf z})$
which states that $z_1^\ast\,F_1({\bf z})$ is a translation-invariant function.
Thus for this representation all equivariant maps have the form
\be
\Ft({\bf z})=\left(\begin{array}{c} F_1({\bf z})\\ F_1(\g 3\cdot{\bf z})\\
F_1(\g 2\cdot{\bf z})\\ F_1({\gt 1\g 2\cdot\bf z})\end{array}\right)
\label{eq:8dmapll}
\ee
where $F_1$ satisfies (\ref{eq:8dref4}) and $(z_1^\ast\,F_1)$ is
$\tor$-invariant. As in the four-dimensional case we determine a Hilbert basis
which generates the $\tor$-invariant functions and then analyze the
equivariants (\ref{eq:8dmapll}) in light of this basis.

\subsubsection{Generators for the $\tor$-invariants}

It is sufficient to enumerate the generators for the monomial invariants, a
problem that has been previously considered by Gomes.\cite{gomes,gomes2}  We
obviously require  the elementary quadratic invariants
\be
\sigma_i=|z_i|^2\hspace{0.5in}i=1,2,3,4;\label{eq:elinv}
\ee
for the analysis of the more complicated invariants it is convenient to define
\be
\omega_i^{\nu_i}\equiv\left\{\begin{array}{c}z_i^{\nu_i}\;\;\mbox{\rm if
$\nu_i\geq0$}\\ {z_i^\ast}^{|\nu_i|}\;\;\mbox{\rm if $\nu_i<0$}
\end{array}\right.
\ee
for any integer $\nu_i$.

\begin{lemma} {\rm (Gomes\cite{gomes,gomes2})} A  $\tor$-invariant monomial
$M({\bf z})=z_1^{\mu_1}{z_1^\ast}^{\mu'_1} z_2^{\mu_2}{z_2^\ast}^{\mu'_2}
z_3^{\mu_3}{z_3^\ast}^{\mu'_3} z_4^{\mu_4} {z_4^\ast}^{\mu'_4}$
can be written in the form
\be
M({\bf z})=\sigma_1^{\nu'_1} \sigma_2^{\nu'_2} \sigma_3^{\nu'_3}
\sigma_4^{\nu'_4}\;m({\bf z})  \label{eq:second}
\ee
where $\nu'_i= \min\; (\mu_i,\mu'_i)$ and
$m({\bf z})=\omega_1^{\nu_1}\omega_2^{\nu_2}\omega_3^{\nu_3}\omega_4^{\nu_4}.$
The exponents $(\nu_1,\nu_2,\nu_3,\nu_4)$ are given by
\beq
\nu_1=-(bl+an)/2&&\nu_2=(bl-an)/2\label{eq:nu1}\\
\nu_3=(al+bn)/2&&\nu_4=(al-bn)/2\label{eq:nu2}
\eeq
where the integers $(a,b)$ are arbitrary except that $(a,b)$ must be even for
$(l+n)$ odd, and for $(l+n)$ even, $(a,b)$ must have the same parity but may be
either even or odd. No more than one exponent $\nu_i$ can be zero.
\end{lemma}
\noindent {\em {\bf Proof}.}
\begin{quote}
\begin{enumerate}
\item The reduction to (\ref{eq:second}) only requires extracting that all
factors of the form $z_i\,z_i^\ast$. The $\tor$-invariance of $M({\bf z})$
implies $m({\bf z})$ must be $\tor$-invariant, this in turn requires
$(\nu_1,\nu_2,\nu_3,\nu_4)$ satisfy
\beq
l\,(\nu_1+\nu_2)+n\,(\nu_3+\nu_4)&=&0\label{eq:t1}\\
n\,(\nu_1-\nu_2)+l\,(\nu_3-\nu_4)&=&0\label{eq:t2}.
\eeq
\item If any two exponents are set to zero, then (\ref{eq:t1}) and
(\ref{eq:t2}) force the remaining two to vanish. Hence no more than one
exponent can be zero. For example, if $\nu_3=\nu_4=0$, then $\nu_1+\nu_2=0$,
and $\nu_1=\nu_2$ hence $\nu_1=\nu_2=0$.
\item Since $l$ and $n$ are assumed relatively prime, the first condition
(\ref{eq:t1}) requires that $(\nu_1+\nu_2)$ contain a factor of $n$ and
$(\nu_3+\nu_4)$ contain a factor of $l$:
$(\nu_1+\nu_2)= -an$ and $(\nu_3+\nu_4)= al;$
similarly the second condition (\ref{eq:t2}) requires
$(\nu_1-\nu_2)= -bl$ and $(\nu_3-\nu_4)= bn.$
\item Solving for $(\nu_1,\nu_2,\nu_3,\nu_4)$ in terms of the integers $(a,b)$
yields (\ref{eq:nu1}) - (\ref{eq:nu2}). Obviously the four numerators on the
right hand side in (\ref{eq:nu1}) - (\ref{eq:nu2}) must be even integers, this
constrains the allowed parities for $(a,b)$. For $(l+n)$ odd, $l$ and $n$ must
have opposite parity so both $a$ and $b$ must be even; for $(l+n)$ even, $l$
and $n$ must be odd and it is only necessary for $(a,b)$ to have the same
parity.
\end{enumerate}
{\bf $\Box$}
\end{quote}

Thus the Hilbert basis contains the elementary quadratic invariants,
and also additional higher order ``mixed'' invariants which generate the
remaining $m({\bf z})$ factor in (\ref{eq:second}). In determining generators
for the mixed invariants, we exploit the fact that $\tor$ is a normal subgroup
of $\semi$, that is, if $\tran{\theta_1}{\theta_2}\in\tor$ then
$\gamma\tran{\theta_1}{\theta_2}\gamma^{-1}\in\tor$ for any $\gamma\in\semi$.
This is obviously true if $\gamma$ is a translation and we can easily check
that it holds for $\gt 1$ and $\g 2$, the generators of $\dt$:
$\gt 1\tran{\theta_1}{\theta_2}\gt 1^{-1}=\tran{-\theta_1}{\theta_2}$ and
$\g 2\tran{\theta_1}{\theta_2}\g 2^{-1}=\tran{\theta_2}{\theta_1}.$
The normality of $\tor$ has the following useful consequence.
\begin{lemma} If $M({\bf z})$ is a $\tor$-invariant function and $\gamma\in\dt$
then $M(\gamma\cdot{\bf z})$ is also a $\tor$-invariant function. In
particular, since $\gt 1\g 2\gt 1\g 2\cdot{\bf z}={\bf z}^\ast$, it follows
that $M({\bf z}^\ast)$ is $\tor$-invariant.
\end{lemma}
\noindent {\em {\bf Proof}.}
\begin{quote} We wish to show that
$ M(\gamma\tran{\theta_1}{\theta_2}\cdot{\bf z})=M(\gamma\cdot{\bf z})$
for arbitrary $({\theta_1},{\theta_2})$. Since $\tor$ is a normal subgroup, for
any $\gamma\in\dt$ and any $\tran{\theta_1}{\theta_2}$ we must have
$\gamma\tran{\theta_1}{\theta_2}\gamma^{-1}= \tran{{\theta'}_1}{{\theta'}_2}$
for some $({{\theta'}_1},{{\theta'}_2})$. Thus we have by direct calculation
\beq
M(\gamma\tran{\theta_1}{\theta_2}\cdot{\bf z})&=&
M(\gamma\tran{\theta_1}{\theta_2}\gamma^{-1}\gamma\cdot{\bf z})=
M(\tran{{\theta'}_1}{{\theta'}_2}\gamma\cdot{\bf z})\nonumber\\
&=&M(\gamma\cdot{\bf z})
\eeq
where the last equation follows from the assumed $\tor$-invariance of $M({\bf
z})$.{\bf $\Box$}
\end{quote}
For example in Table 2 we show the $\dt$ orbit of the monomial $m({\bf z})=
{z_1^\ast}^\alpha\,z_3^\delta\,z_4^\sigma$. If the exponents
$\alpha,\delta,\sigma$ are chosen so that $m({\bf z})$ is $\tor$-invariant,
then by Lemma II.3 each monomial on the orbit of $m({\bf z})$ is also
$\tor$-invariant.

\begin{table}\centering
\begin{center}
Table 2.  $\dt$ acting on ${E_{\k^2}(\ext)}$ and the $\dt$-orbit of $m({\bf
z})={z_1^\ast}^\alpha\,z_3^\delta\,z_4^\sigma$
\end{center}
\vspace{7mm}
\begin{tabular}{cccc}
\underline{$\gamma\in\dt$}
&\underline{$\gamma\cdot{\bf z}$} & \underline{$m(\gamma\cdot{\bf z})$}&
\underline{$(\nu_1,\nu_2,\nu_3,\nu_4)$}\\
\\
$I$ & $(z_1,z_2,z_3,z_4)$ &
${z_1^\ast}^\alpha\,z_3^\delta\,z_4^\sigma$ & $(-\alpha,0,\delta,\sigma)$\\
\\
$\gt 1$ &  $(z_2^\ast,z_1^\ast,z_4^\ast,z_3^\ast)$ &
${z_2}^\alpha\,{z_3^\ast}^\sigma\,{z_4^\ast}^\delta$&
$(0,\alpha,-\sigma,-\delta)$\\
\\
$\g 2$ &  $(z_3,z_4^\ast,z_1,z_2^\ast)$ &
${z_1}^\delta\,{z_2^\ast}^\sigma\,{z_3^\ast}^\alpha$&
$(\delta,-\sigma,-\alpha,0)$\\
\\
$\gt 1\g 2$ &  $(z_4,z_3^\ast,z_2,z_1^\ast)$&
${z_1^\ast}^\sigma\,{z_2}^\delta\,{z_4^\ast}^\alpha$&
$(-\sigma,\delta,0,\alpha)$\\
\\
$\g 3$ &  $(z_2,z_1,z_4,z_3)$ &
${z_2^\ast}^\alpha\,{z_3}^\sigma\,{z_4}^\delta$& $(0,-\alpha,\sigma,\delta)$\\
\\
$\gt 1\g 3$ &  $(z_1^\ast,z_2^\ast,z_3^\ast,z_4^\ast)$ &
${z_1}^\alpha\,{z_3^\ast}^\delta\,{z_4^\ast}^\sigma$&
$(\alpha,0,-\delta,-\sigma)$\\
\\
$\g 2\gt 1$ &  $(z_4^\ast,z_3,z_2^\ast,z_1)$ &
${z_1}^\sigma\,{z_2^\ast}^\delta\,{z_4}^\alpha$& $(\sigma,-\delta,0,\alpha)$\\
\\
$\gt 1\g 2\gt 1$ &  $(z_3^\ast,z_4,z_1^\ast,z_2)$&
${z_1^\ast}^\delta\,{z_2}^\sigma\,{z_3}^\alpha$& $(-\delta,\sigma,\alpha,0)$\\
\end{tabular}
\par
\vspace{7 mm}
\underline{Notes:}\raggedright
\par
Here ${\bf z}=(z_1,z_2,z_3,z_4)$ denotes a point in ${E_{\k^2}(\ext)}$, $I$
stands for the identity element, and $\g 3\equiv\g 2\gt 1\g 2$. The exponents
$(\alpha,\delta,\sigma)$ are positive integers.
\protect \vspace*{\fill}
\end{table}
\begin{table}[p]\centering
\begin{center}
Table 3.  $\dt$ acting on ${E_{\k^2}(\ext)}$ and the $\dt$-orbits of monomials
$m({\bf z})=\omega_1^{\nu_1}\omega_2^{\nu_2}\omega_3^{\nu_3}\omega_4^{\nu_4}$
\end{center}
\vspace{7mm}
\begin{tabular}{ccccc}
\underline{$\gamma\in\dt$}&
\underline{$m_1(\gamma\cdot{\bf z})$}& \underline{$m_2(\gamma\cdot{\bf z})$}&
\underline{$m_3(\gamma\cdot{\bf z})$} &\underline{$m_4(\gamma\cdot{\bf z})$}\\
\\
$I$ &  ${\bf{z_1^\ast}^\alpha\,z_2^\beta\,z_3^\delta\,z_4^\sigma}$ &
${\bf{z_1^\ast}^\alpha\,{z_2^\ast}^\beta\,{z_3}^\delta\,{z_4}^\sigma}$ &
${\bf{z_1^\ast}^\alpha\,{z_2^\ast}^\beta\,{z_3^\ast}^\delta\,{z_4}^\sigma}$ &
${\bf{z_1^\ast}^\alpha\,{z_2^\ast}^\beta\,{z_3^\ast}^\delta\,{z_4^\ast}^\sigma}$\\
\\
$\gt 1$&
${\bf{z_1^\ast}^\beta\,{z_2}^\alpha\,{z_3^\ast}^\sigma\,{z_4^\ast}^\delta}$&
${\bf{z_1}^\beta\,{z_2}^\alpha\,{z_3^\ast}^\sigma\,{z_4^\ast}^\delta}$&
${\bf{z_1}^\beta\,{z_2}^\alpha\,{z_3^\ast}^\sigma\,{z_4}^\delta}$&
${\bf{z_1}^\beta\,{z_2}^\alpha\,{z_3}^\sigma\,{z_4}^\delta}$\\
\\
$\g 2$&
${\bf{z_1}^\delta\,{z_2^\ast}^\sigma\,{z_3^\ast}^\alpha\,{z_4^\ast}^\beta}$&
${\bf{z_1}^\delta\,{z_2^\ast}^\sigma\,{z_3^\ast}^\alpha\,{z_4}^\beta}$&
${z_1^\ast}^\delta\,{z_2^\ast}^\sigma\,{z_3^\ast}^\alpha\,{z_4}^\beta$&
${\bf{z_1^\ast}^\delta\,{z_2}^\sigma\,{z_3^\ast}^\alpha\,{z_4}^\beta}$\\
\\
$\gt 1\g 2$&
${z_1^\ast}^\sigma\,{z_2}^\delta\,{z_3^\ast}^\beta\,{z_4^\ast}^\alpha$&
${\bf{z_1^\ast}^\sigma\,{z_2}^\delta\,{z_3}^\beta\,{z_4^\ast}^\alpha}$&
${\bf{z_1^\ast}^\sigma\,{z_2^\ast}^\delta\,{z_3}^\beta\,{z_4^\ast}^\alpha}$&
${\bf{z_1}^\sigma\,{z_2^\ast}^\delta\,{z_3}^\beta\,{z_4^\ast}^\alpha}$\\
\\
$\g 3$ & ${\bf{z_1}^\beta\,{z_2^\ast}^\alpha\,{z_3}^\sigma\,{z_4}^\delta}$&
${z_1^\ast}^\beta\,{z_2^\ast}^\alpha\,{z_3}^\sigma\,{z_4}^\delta$&
${z_1^\ast}^\beta\,{z_2^\ast}^\alpha\,{z_3}^\sigma\,{z_4^\ast}^\delta$&
${z_1^\ast}^\beta\,{z_2^\ast}^\alpha\,{z_3^\ast}^\sigma\,{z_4^\ast}^\delta$\\
\\
$\gt 1\g 3$ &
${z_1}^\alpha\,{z_2^\ast}^\beta\,{z_3^\ast}^\delta\,{z_4^\ast}^\sigma$&
${z_1}^\alpha\,{z_2}^\beta\,{z_3^\ast}^\delta\,{z_4^\ast}^\sigma$&
${\bf{z_1}^\alpha\,{z_2}^\beta\,{z_3}^\delta\,{z_4^\ast}^\sigma}$&
${z_1}^\alpha\,{z_2}^\beta\,{z_3}^\delta\,{z_4}^\sigma$\\
\\
$\g 2\gt 1$ &${z_1}^\sigma\,{z_2^\ast}^\delta\,{z_3}^\beta\,{z_4}^\alpha$&
${z_1}^\sigma\,{z_2^\ast}^\delta\,{z_3^\ast}^\beta\,{z_4}^\alpha$&
${z_1}^\sigma\,{z_2}^\delta\,{z_3^\ast}^\beta\,{z_4}^\alpha$&
${z_1^\ast}^\sigma\,{z_2}^\delta\,{z_3^\ast}^\beta\,{z_4}^\alpha$\\
\\
$\gt 1\g 2\gt 1$ &${z_1^\ast}^\delta\,{z_2}^\sigma\,{z_3}^\alpha\,{z_4}^\beta$&
${z_1^\ast}^\delta\,{z_2}^\sigma\,{z_3}^\alpha\,{z_4^\ast}^\beta$&
${z_1}^\delta\,{z_2}^\sigma\,{z_3}^\alpha\,{z_4^\ast}^\beta$&
${z_1}^\delta\,{z_2^\ast}^\sigma\,{z_3}^\alpha\,{z_4^\ast}^\beta$\\
\end{tabular}
\par
\vspace{7 mm}
\underline{Notes:}\raggedright
\par
Here ${\bf z}=(z_1,z_2,z_3,z_4)$ denotes a point in ${E_{\k^2}(\ext)}$, $I$
stands for the identity element, and $\g 3\equiv\g 2\gt 1\g 2$. The exponents
$(\alpha,\beta,\delta,\sigma)$ are positive integers. There are $2^4=16$
distinct ways to choose the signs of the exponents $(\nu_1,\nu_2,\nu_3,\nu_4)$;
each possibility occurs twice in the table. The boldfaced entries comprise one
set of the $16$ possibilities.
\protect \vspace*{\fill}
\end{table}

In light of Lemma II.3, we seek a {\em sub-basis} of $\tor$ invariants
$\{\sigma_i\}_{i=1}^J$ with the property that collectively the $\dt$ orbits of
the sub-basis provide a Hilbert basis for the $\tor$-invariants.\footnote{A
different strategy for calculating translation invariants has been described by
Gomes as part of an investigation of hidden symmetry in bifurcation problems
with rectangular symmetry.\cite{gomes}} In enumerating this sub-basis we have
the elementary invariant $\sigma_1=|z_1|^2$, whose $\dt$ orbit consists of the
four quadratic invariants (\ref{eq:elinv}), and a variable set of mixed
invariants whose number and definition depends on the representation $(l,n)$.
Three of these can be given explicitly:
\beq
\sigma_5({\bf z})&=&\left\{\begin{array}{cc}
{z_1^\ast}^{nl}\;z_3^{(l^2+n^2)/2}\;z_4^{(l^2-n^2)/2}&\mbox{\rm if  $(l+n)$
even}\\
&\\
&\\
{z_1^\ast}^{2nl}\;z_3^{(l^2+n^2)}\;z_4^{(l^2-n^2)}&\mbox{\rm if $(l+n)$ odd}
\end{array}\right.\label{eq:s5}\\
&&\nonumber\\
&&\nonumber\\
\sigma_6({\bf z})&=&\left\{\begin{array}{cc}
(z_1^\ast)^{(l+n)/2}z_2^{(l-n)/2}z_3^{(l+n)/2}z_4^{(l-n)/2}&\mbox{\rm if
$(l+n)$ even}\\
&\\
&\\
(z_1^\ast)^{(l+n)}z_2^{(l-n)}z_3^{(l+n)}z_4^{(l-n)}&\mbox{\rm if $(l+n)$ odd}
\end{array}\right.\label{eq:s6}\\
&&\nonumber\\
\sigma_7({\bf z})&=&(z_1^\ast z_2^\ast)^n\;(z_3 z_4)^l.\label{eq:s7}
\eeq
In addition the sub-basis generally involves invariants that are most simply
described in a somewhat implicit fashion. Given a positive integer $\beta$,
define the integers $(a_\beta,b_\beta)$ and $({a'}_\beta ,{b'}_\beta )$ as
follows. Let $a_\beta$ denote the smallest positive integer (for fixed $\beta$)
such that
$b_\beta\equiv({2\beta+ n a_\beta})/{l}$
is an integer with the same even/odd parity as $a_\beta$. Similarly let
${b'}_\beta$ denote the smallest positive integer (for fixed $\beta$) such that
${a'}_\beta \equiv({2\beta+ l {b'}_\beta})/{n}$
is an integer with the same even/odd parity as ${b'}_\beta $. Using
$(a_\beta,b_\beta)$ and $({a'}_\beta ,{b'}_\beta )$ we define the two
$\tor$-invariants:
\beq
m_1({\bf z},a_\beta,b_\beta)&\equiv& (z_1^\ast)^{(b_\beta l+ a_\beta
n)/2}z_2^{(b_\beta l-a_\beta n)/2}z_3^{(a_\beta l+b_\beta n)/2}z_4^{(a_\beta
l-b_\beta n)/2}.\label{eq:m1min}\\
m_2({\bf z},{a'}_\beta ,{b'}_\beta )&\equiv& (z_1^\ast)^{({b'}_\beta  l+
{a'}_\beta  n)/2}(z_2^\ast)^{({a'}_\beta  n-{b'}_\beta  l)/2}z_3^{({a'}_\beta
l+{b'}_\beta  n)/2}z_4^{({a'}_\beta l-{b'}_\beta  n)/2}.\label{eq:m2min}
\eeq

\begin{lemma} For the eight-dimensional representation {\rm (\ref{eq:8drepa}) -
(\ref{eq:8drepc})} with $l>n>0$ relatively prime, the invariants
$\{\sigma_i\}_{i=1}^7$, $\{m_1({\bf z},a_\beta,b_\beta)\}_{\beta=1}^{N}$, and
$\{m_2({\bf z},{a'}_\beta ,{b'}_\beta )\}_{\beta=1}^{n-1}$ where
\be
N=\left\{\begin{array}{cc}
(l-n-2)/2&\mbox{\rm if  $(l+n)$ even}\\
&\\
l-n-1&\mbox{\rm if $(l+n)$ odd}.
\end{array}\right.\label{eq:Ndef}
\ee
form a sub-basis for the $\tor$-invariants. If $N=0$ the $m_1$ invariants are
absent and if $n=1$ the $m_2$ invariants are absent.
\end{lemma}
\noindent {\em {\bf Proof}. }
\begin{quote} This result follows from Lemma II.2 and Lemmas II.5, II.6 below.
{\bf $\Box$}
\end{quote}

As an illustration of Lemma II.4 let $(l,n)=(4,3)$, then we have $N=0$ in
(\ref{eq:Ndef}) so there are no invariants of $m_1$ type. There are two
invariants of $m_2$ type corresponding to $\beta=1,2$ with $({a'}_1
,{b'}_1)=(6,4)$ and $({a'}_2 ,{b'}_2)=(4,2)$ respectively. The sub-basis
consists of the quadratic invariants $\sigma_1, \ldots,\sigma_4$ and
\beq
\sigma_5&=&{z_1^\ast}^{24}\;z_3^{25}\;z_4^{7}\nonumber\\
\sigma_6&=&(z_1^\ast)^{7}z_2\,z_3^{7}z_4\nonumber\\
\sigma_7&=&(z_1^\ast z_2^\ast)^3\;(z_3 z_4)^4\\
\sigma_8&=&m_2({\bf z},6,4)=
(z_1^\ast)^{17}\,z_2^\ast\,z_3^{18}z_4^{6}\nonumber\\
\sigma_9&=&m_2({\bf z},4,2)=
(z_1^\ast)^{10}(z_2^\ast)^{2}\,z_3^{11}z_4^{5}.\nonumber
\eeq
When $\dt$ is applied to this sub-basis, we obtain a Hilbert basis with $36$
distinct elements.

We now give two results that characterize a set of generators for the mixed
invariants.

\begin{lemma} \begin{enumerate}
\item All the $\tor$-invariant monomials
$m({\bf z})=\omega_1^{\nu_1}\omega_2^{\nu_2}\omega_3^{\nu_3}\omega_4^{\nu_4}$
with exactly one zero exponent $\nu_i=0$ are generated by applying $\dt$ to the
invariant $\sigma_5({\bf z})$.
\item  All the $\tor$-invariant monomials $m({\bf
z})=\omega_1^{\nu_1}\omega_2^{\nu_2}\omega_3^{\nu_3}\omega_4^{\nu_4}$ with
$\nu_i\neq0$ for $i=1,2,3,4$ are generated by applying $\dt$ to the following
two classes of monomial invariants:
\begin{enumerate}
\item Class $m_1$:
\be
m_1({\bf z},a,b)=
{z_1^\ast}^\alpha\,z_2^\beta\,z_3^\delta\,z_4^\sigma\label{eq:m1}
\ee
with
\beq
\alpha=(bl+an)/2&&\beta=(bl-an)/2\label{eq:m1t1}\\
\delta=(al+bn)/2&&\sigma=(al-bn)/2\label{eq:m1t4}
\eeq
where the positive integers $(a,b)$ satisfy
\be
a\geq b>\frac{n}{l}a>0.\label{eq:estf}
\ee

\item Class $m_2$:
\be
m_2({\bf z},a,b)=
{z_1^\ast}^\alpha\,{z_2^\ast}^\beta\,{z_3}^\delta\,{z_4}^\sigma \label{eq:m2}
\ee
with
\beq
\alpha=(bl+an)/2&&\beta=(an-bl)/2\label{eq:m2t1}\\
\delta=(al+bn)/2&&\sigma=(al-bn)/2\label{eq:m2t4}
\eeq
where the non-negative integers $(a,b)$ satisfy
\be
\frac{n}{l}a> b\geq 0.\label{eq:estff}
\ee
\end{enumerate}
\end{enumerate}
\end{lemma}
\noindent {\em {\bf Proof}.}
\begin{quote}
\begin{enumerate}
\item  First consider monomials $m({\bf z})$ with exactly one zero exponent.
{}From Table 2 it is clear that the $\dt$ orbit of every such monomial must
contain a point with $\nu_2=0$ (in fact two such points). Thus we can restrict
attention to invariants $m({\bf
z})=\omega_1^{\nu_1}\omega_3^{\nu_3}\omega_4^{\nu_4}$ with $\nu_2=0$ and then
apply $\dt$ to generate the remaining possibilities. In addition we can always
arrange that $\nu_1<0$ by applying $\g 2\gt 1$ if necessary. From
(\ref{eq:nu1}) - (\ref{eq:nu2}), setting $\nu_2=0$ requires $bl=an$ which in
turn requires that $b$ contain a factor of $n$ and $a$ a factor of $l$:
$a=dl$ and $b=dn.$
Hence the remaining exponents become
$\nu_1=-dnl$, $\nu_3=d(l^2+n^2)/2$, and $\nu_4=d(l^2-n^2)/2;$
$d<0$ is required to obtain $\nu_1<0$. Thus all of these invariants have the
form
\be
\left[(z_1^\ast)^{nl}\,z_3^{(l^2+n^2)/2}\,z_4^{(l^2-n^2)/2}\right]^d;
\label{eq:sig5}
\ee
if $(l+n)$ is even then $d=1$ gives the lowest order result and if $(l+n)$ is
odd then we must take $d=2$. This yields the definition of $\sigma_5$ in
(\ref{eq:s5}), and proves the first part of the lemma.

\item Now consider monomials $m({\bf z})
=\omega_1^{\nu_1}\omega_2^{\nu_2}\omega_3^{\nu_3}\omega_4^{\nu_4}$ with all
exponents non-zero. There are $2^4=16$ ways to choose the signs of
$(\nu_1,\nu_2,\nu_3,\nu_4)$, however we need only consider the four
possibilites:
$m_1({\bf z})={z_1^\ast}^\alpha\,z_2^\beta\,z_3^\delta\,z_4^\sigma$,
$m_2({\bf z})={z_1^\ast}^\alpha\,{z_2^\ast}^\beta\,{z_3}^\delta\,{z_4}^\sigma$,
$m_3({\bf z})=
{z_1^\ast}^\alpha\,{z_2^\ast}^\beta\,{z_3^\ast}^\delta\,{z_4}^\sigma$, and
$m_4({\bf z})=
{z_1^\ast}^\alpha\,{z_2^\ast}^\beta\,{z_3^\ast}^\delta\,{z_4^\ast}^\sigma$;
the remaining twelve possibilities can be generated by applying $\dt$, cf.
Table 3.

\item Of these four possibilities, only the first two can satisfy the
requirement of $\tor$-invariance. For $m_3$, the relations (\ref{eq:nu1}) -
(\ref{eq:nu2}) become
$\alpha=(bl+an)/2$, $\beta=-(bl-an)/2$, $\delta=-(al+bn)/2$, and
$\sigma=(al-bn)/2.$
Therefore $\alpha+\beta=an$ and $\delta+\sigma=-bl$ imply
$a>0$ and $b<0,$
so that  $\alpha>0$ and $\delta>0$ require $|a|>|b|$ and $|a|<|b|$,
respectively. This is impossible.
Similarly for $m_4$, $\tor$-invariance requires
$\alpha=(bl+an)/2\label{eq:m41}$, $\beta=-(bl-an)/2$, $\delta=-(al+bn)/2$, and
$\sigma=-(al-bn)/2;$
now $\alpha+\beta=an$ and $\delta+\sigma=-al$ require $a>0$ and $a<0$
respectively. Thus the monomials of type $m_3$ and $m_4$ cannot be
$\tor$-invariant.

\item For the $m_1$ monomials, translation invariance requires (\ref{eq:m1t1})
- (\ref{eq:m1t4}). In this case $\alpha+\beta=bl$ and $\delta+\sigma=al$ imply
$a>0$ and $b>0$, respectively. In conjunction with  $\beta>0$ and $\sigma>0$,
this yields
\be
\frac{al}{n}>b>\frac{an}{l}>0;\label{eq:esti}
\ee
finally applying $\gt 1\g 2\gt 1$ interchanges the roles of $a$ and $b$ so we
can replace (\ref{eq:esti}) with (\ref{eq:estf}).

\item  For the $m_2$ monomials, translation invariance requires (\ref{eq:m2t1})
- (\ref{eq:m2t4}). Now $\alpha+\beta=an$ implies $a>0$, and $\beta>0$ requires
${an}/{l}>b.$
Although the sign of $b$ is not determined, by applying $\g 3$ we can arrange
for $b$ to be non-negative. These requirements are summarized in
(\ref{eq:estff}). This proves the second part of the lemma. {\bf $\Box$}
\end{enumerate}
\end{quote}

We now describe how to generate the remaining two classes $m_1$ and $m_2$ from
a finite number of invariants. The answer depends on whether $(l+n)$ is odd or
even, and involves the invariants defined in (\ref{eq:m1min}) and
(\ref{eq:m2min}).

\begin{lemma}
\begin{enumerate}

\item For $(l+n)$ even:
\begin{enumerate}
\item The $\tor$-invariant monomials of the form $m_1$ in {\rm (\ref{eq:m1})}
can be generated by $\sigma_5$, $\sigma_6$
and $m_1({\bf z},a_\beta,b_\beta)$ for $\beta=1,2,\ldots,(l-n-2)/2.$

\item The $\tor$-invariant monomials of the form $m_2$ in {\rm (\ref{eq:m2})}
can be generated by $\sigma_5$, $\sigma_7$, and $m_2({\bf z},{a'}_\beta
,{b'}_\beta )$ for $\beta=1,2,\ldots,n-1$.

\end{enumerate}

\item For $(l+n)$ odd:
\begin{enumerate}
\item The $\tor$-invariant monomials of the form $m_1$ in {\rm (\ref{eq:m1})}
can be generated by $\sigma_5$, $\sigma_6$,
and $m_1({\bf z},a_\beta,b_\beta)$ for $\beta=1,2,\ldots,(l-n)-1.$

\item The $\tor$-invariant monomials of the form $m_2$ in {\rm (\ref{eq:m2})}
can be generated by $\sigma_5$, $\sigma_7$, and $m_2({\bf z},{a'}_\beta
,{b'}_\beta )$ for $\beta=1,2,\ldots,n-1$.

\end{enumerate}
\end{enumerate}

\end{lemma}
\noindent {\em {\bf Proof}.}
\begin{quote}
We analyze the classes $m_1$ and $m_2$ separately.
\begin{enumerate}
\item \underline{Invariants of $m_1$ form.}
\begin{enumerate}

\item Assume $(l+n)$ is odd; in this case $(a,b)$ must be even. Given $m_1({\bf
z},a,b)$, if $\beta=(bl-an)/2\geq l-n$ then we can extract a factor of
$\sigma_6$: $m_1({\bf z},a,b)=\sigma_6\;m_1({\bf z},a-2,b-2),$
and reduce $\beta$ by $(l-n)$. Thus we need only consider invariants $m_1({\bf
z},a,b)$ such that $\beta<l-n$. All of these can be written as
$m_1({\bf z},a,b)={\sigma_5}^k\;m_1({\bf z},a_\beta,b_\beta)$
for some integer $k\geq0$. Hence we require $\sigma_5$, $\sigma_6$, and the
invariants $m_1({\bf z},a_\beta,b_\beta)$ for $\beta=1,2,\ldots,(l-n)-1.$

\item Assume $(l+n)$ is even, then $(l,n)$ are each odd and $(a,b)$ may be
either even or odd. Given $m_1({\bf z},a,b)$, if $\beta=(bl-an)/2\geq (l-n)/2$
then either $a=b=1$ or $a\geq b>1$. In the former case
$m_1({\bf z},1,1)=\sigma_6;$
in the latter case we can extract a factor of $\sigma_6$:
$m_1({\bf z},a,b)=\sigma_6\;m_1({\bf z},a-1,b-1),$
and reduce $\beta$ by $(l-n)/2$. Thus we need only consider invariants
$m_1({\bf z},a,b)$ such that $\beta<(l-n)/2$. All of these can be written as
$m_1({\bf z},a,b)={\sigma_5}^k\;m_1({\bf z},a_\beta,b_\beta)$
for some integer $k\geq0$. Hence we require $\sigma_5$, $\sigma_6$, and the
invariants $m_1({\bf z},a_\beta,b_\beta)$ for $\beta=1,2,\ldots,(l-n-2)/2.$

\end{enumerate}
\item \underline{Invariants of $m_2$ form.} Given $m_2({\bf z},a,b)$, if
$\beta=(an-bl)/2\geq n$ then we can extract a factor of $\sigma_7$:
$ m_2({\bf z},a,b)=\sigma_7\;m_2({\bf z},a-2,b),$
and reduce $\beta$ by $n$. Thus we need only consider invariants $m_2({\bf
z},a,b)$ such that $\beta<n$. All of these can be written as
$m_2({\bf z},a,b)={\sigma_5}^k\;m_2({\bf z},{a'}_\beta ,{b'}_\beta )$
for some integer $k\geq0$. Hence we require $\sigma_5$, $\sigma_7$, and the
invariants $m_2({\bf z},{a'}_\beta ,{b'}_\beta )$ for $\beta=1,2,\ldots,(n-1).$
\end{enumerate}
{\bf $\Box$}
\end{quote}
This completes the proof of Lemma II.4 enumerating a sub-basis for the
$\tor$-invariants.

\subsubsection{Extension from ${E_{\k^2}(\ext)}$ to $E_{\k^2}(\rtwo)$}

We return to the main issue: whether an arbitrary $\semi$ symmetric map can be
obtained by restriction from an $\euc$-equivariant map.

\begin{lemma} Assume ${E_{\k^2}(\ext)}$ carries an eight-dimensional
irreducible representation of $\semi$ as in {\rm (\ref{eq:8drepa}) -
(\ref{eq:8drepc})}. Let $\Ft:{E_{\k^2}(\ext)}\rightarrow{E_{\k^2}(\ext)}$
denote a $\semi$-equivariant map, then there is a $\euc$-equivariant map on
$E_{\k^2}(\rtwo)$, $\strobe:{E_{\k^2}(\rtwo)}\rightarrow{E_{\k^2}(\rtwo)},$
such that
\be
\Ft = \strobe |_{{E_{\k^2}(\ext)}}.\label{eq:8dpbc2}
\ee
\end{lemma}
\noindent {\em {\bf Proof}.}
\begin{quote}
\begin{enumerate}
\item The argument follows the proof of lemma II.1. Recall from
(\ref{eq:8dmapll}) that $\Ft$ has the form
\be
\Ft({\bf z})=\left(\begin{array}{c} F_1({\bf z})\\ F_1(\g 3\cdot{\bf z})\\
F_1(\g 2\cdot{\bf z})\\ F_1({\gt 1\g 2\cdot\bf z})\end{array}\right)
\label{eq:8dmapf1}
\ee
where $F_1$ satisfies $F_1({\bf z})=F_1({\bf z}^\ast)^\ast$ and
$z_1^\ast\,F_1({\bf z})$ is $\tor$ invariant. Without loss of generality we
assume $z_1^\ast\,F_1({\bf z})$ is a $\tor$-invariant monomial, then
\be
z_1^\ast\,F_1({\bf z})= \sigma_1^{\nu'_1} \sigma_2^{\nu'_2} \sigma_3^{\nu'_3}
\sigma_4^{\nu'_4}
\omega_1^{\nu_1}\omega_2^{\nu_2}\omega_3^{\nu_3}\omega_4^{\nu_4}
\label{eq:tmon}
\ee
from (\ref{eq:second}). Any $\tor$-invariant monomial can appear on the right
hand side provided it contains a factor of $z_1^\ast$.

\item On the other hand, when we restrict a $\euc$-symmetric map on
${E_{\k^2}(\rtwo)}$ to the eigenspace ${E_{\k^2}(\ext)}$ we get a
$\semi$-symmetric map with $F_1$ of the form
\be
F_1(z_1,z_2)=\sum_{\vec{k'_1}\in \tilde{A}(\k)}\ldots\sum_{\vec{k'_p}\in
\tilde{A}(\k)} a(\vec{k'_1})\,a(\vec{k'_2})\,\ldots
a(\vec{k'_p})\,P(\vec{k_1},\vec{k'_1},\vec{k'_2}, \ldots,\vec{k'_p})
\label{eq:F1nf8d}
\ee
where
\beq
a(\vec{k})&=&z_1\delta_{\vec{k},\vec{k_1}}+z_2\delta_{\vec{k},\vec{k_2}} +
z_3\delta_{\vec{k},\vec{k_3}}+z_4\delta_{\vec{k},\vec{k_4}}\nonumber\\
&& +z_1^\ast\delta_{\vec{k},\vec{-k_1}}+z_2^\ast\delta_{\vec{k},\vec{-k_2}} +
+z_3^\ast\delta_{\vec{k},\vec{-k_3}}+z_4^\ast\delta_{\vec{k},\vec{-k_4}}
\label{eq:restrict8d}
\eeq
and $\tilde{A}(\k)=\{\pm\vec{k_1},\pm\vec{k_2},\pm\vec{k_3},\pm\vec{k_4}\}$
from (\ref{eq:akint}). The function $P(\vec{k_1},\vec{k'_1},\vec{k'_2},
\ldots,\vec{k'_p})$ satisfies the conditions (\ref{eq:cond1}) -
(\ref{eq:cond2}). Thus the extension of the $\semi$-symmetric map
(\ref{eq:8dmapf1}) to a $\euc$-symmetric map depends on the existence of
$P(\vec{k_1},\vec{k'_1},\vec{k'_2}, \ldots,\vec{k'_p})$ such that
\be
\frac{\sigma_1^{\nu'_1} \sigma_2^{\nu'_2} \sigma_3^{\nu'_3} \sigma_4^{\nu'_4}
\omega_1^{\nu_1}\omega_2^{\nu_2}\omega_3^{\nu_3}\omega_4^{\nu_4}}{z_1^\ast} =
\sum_{\vec{k'_1}\in \tilde{A}(\k)}\ldots\sum_{\vec{k'_p}\in \tilde{A}(\k)}
a(\vec{k'_1})\,a(\vec{k'_2})\,\ldots
a(\vec{k'_p})\,P(\vec{k_1},\vec{k'_1},\vec{k'_2}, \ldots,\vec{k'_p}).
\label{eq:key}
\ee
We shall check that $P$ can always be found, given any appropriate
$\tor$-invariant monomial in (\ref{eq:tmon}).

\item The $O(2)$ invariance of $P$ will be assured if it is constructed from
the inner products of its arguments. For the eight-dimensional representation,
we have
\be
\left(\begin{array}{cccc}
\vec{k_1}\cdot\vec{k_1}&\vec{k_1}\cdot\vec{k_2}&\vec{k_1}\cdot\vec{k_3}
&\vec{k_1}\cdot\vec{k_4}\\
\\
\vec{k_2}\cdot\vec{k_1}&\vec{k_2}\cdot\vec{k_2}&\vec{k_2}\cdot\vec{k_3}
&\vec{k_2}\cdot\vec{k_4}\\
\\
\vec{k_3}\cdot\vec{k_1}&\vec{k_3}\cdot\vec{k_2}&\vec{k_3}\cdot\vec{k_3}
&\vec{k_3}\cdot\vec{k_4}\\
\\
\vec{k_4}\cdot\vec{k_1}&\vec{k_4}\cdot\vec{k_2}&\vec{k_4}\cdot\vec{k_3}
&\vec{k_4}\cdot\vec{k_4}
\end{array}\right)=\left(\begin{array}{cccc}
\k^2&l^2-n^2&2nl&0\\
\\
l^2-n^2&\k^2&0&2nl\\
\\
2nl&0&\k^2&n^2-l^2\\
\\
0&2nl&n^2-l^2&\k^2
\end{array}\right)\label{eq:matrix8d}
\ee
so that for vectors in $\tilde{A}(\k)$ the inner product
$\vec{k_i}\cdot\vec{k_j}$ has only four possible values, up to an overall sign.
Define
$I_1(\vec{k},\vec{q})=\delta_{\vec{k}\cdot \vec{q},\k^2}$,
$I_2(\vec{k},\vec{q})=\delta_{\vec{k}\cdot \vec{q},l^2-n^2}$,
$I_3(\vec{k},\vec{q})=\delta_{\vec{k}\cdot \vec{q},2nl}$, and
$I_4(\vec{k},\vec{q})=\delta_{\vec{k}\cdot \vec{q},0}$
to describe these four values.

\item The monomial at left in (\ref{eq:key}) contains a finite number of
factors of the coordinates $(z_1,z_2,z_3,z_4)$ and their complex conjugates. If
each of these eight coordinates can be individually specified using the
functions $I_i$, then we obviously can construct an $O(2)$-invariant $P$ to
yield a given monomial. Note, that in (\ref{eq:key}) $\vec{k_1}$ is always the
first argument in $P$ regardless of which monomial we are given; this allows us
to specify factors of $z_1$, $z_2$, and $z_3$ using $I_1$, $I_2$, and $I_3$,
respectively:
\be
\sum_{\vec{k'_1}\in \tilde{A}(\k)} a(\vec{k'_1})\,I_i(\vec{k_1},\vec{k'_1})=
z_i\;\;\;\;\;\;\mbox{for}\;\;i=1,2,3;\label{eq:p1lin}
\ee
similarly using $I_i(\vec{k_1},-\vec{k'_1})$ yields $z_i^\ast$. Thus arbitrary
monomials in $(z_1,z_2,z_3,z_1^\ast,z_2^\ast,z_3^\ast)$ may be obtained from
$O(2)$-invariant $P$.

\item Unfortunately $z_4$ and $z_4^\ast$ are not uniquely specified by $I_4$ in
this way, instead we obtain
\be
\sum_{\vec{k'_1}\in \tilde{A}(\k)} a(\vec{k'_1})\,I_4(\vec{k_1},\vec{k'_1})=
z_4+z_4^\ast
\ee
since $+\vec{k_4}$ and $-\vec{k_4}$ are both perpendicular to $\vec{k_1}$.
However our analysis of the $\tor$-invariant monomials (\ref{eq:tmon}) shows
that $z_4$ and $z_4^\ast$ appear either together in $\sigma_4=|z_4|^2$ or
separately in one of the higher order mixed invariants. Factors of $\sigma_4$
can be easily specified with $I_1$:
\be
\sum_{\vec{k'_1}\in \tilde{A}(\k)} \sum_{\vec{k'_2}\in \tilde{A}(\k)}
a(\vec{k'_1})\,a(\vec{k'_2})\,
\frac{I_4(\vec{k_1},\vec{k'_1})\,I_1(\vec{k'_1},-\vec{k'_2})}{2} = \sigma_4.
\label{eq:sig4}
\ee
When $z_4$ or $z_4^\ast$ appears in (\ref{eq:tmon}) through one of the higher
order invariants then it will always be in combination with at least two of the
other three coordinates $(z_1,z_2,z_3)$ or their complex conjugates. Inspection
of the matrix of inner product values (\ref{eq:matrix8d}) shows that the
combination of $z_4$ or $z_4^\ast$ with any one of the other three coordinates
can be uniquely specified in terms of the $I_i$. For example consider
$z_2^\ast\,z_4$, since $\vec{k_1}\cdot\vec{k_2}=\k^2$ and
$\vec{k_2}\cdot\vec{k_4}=2nl$ we use
$I_2(\vec{k_1},-\vec{k'_1})\,I_3(-\vec{k'_1},\vec{k'_2})$:
\be
\sum_{\vec{k'_1}\in \tilde{A}(\k)} \sum_{\vec{k'_2}\in \tilde{A}(\k)}
a(\vec{k'_1})\,a(\vec{k'_2})\,
I_2(\vec{k_1},-\vec{k'_1})\,I_3(-\vec{k'_1},\vec{k'_2}) = z_2^\ast\,z_4.
\ee
Thus in (\ref{eq:key}) given any monomial on the left, we can construct an
$O(2)$-invariant function $P$ which yields that monomial.

\item It remains to verify that these $O(2)$-invariant functions may be
constructed so as to also satisfy the requirement (\ref{eq:cond1}) of
translation symmetry:
\be
P(\vec{k},\vec{k'_1},\vec{k'_2},\ldots,\vec{k'_p})=0\;\;\mbox{\rm unless}\;\;
\vec{k}=\vec{k'_1}+\vec{k'_2}+\ldots+\vec{k'_p}.\label{eq:c1}
\ee
It is helpful to formalize the foregoing construction somewhat.  Any given
monomial in (\ref{eq:key}),
\be
F_1(z_1,z_2)=\frac{\sigma_1^{\nu'_1} \sigma_2^{\nu'_2} \sigma_3^{\nu'_3}
\sigma_4^{\nu'_4}\omega_1^{\nu_1}\omega_2^{\nu_2}\omega_3^{\nu_3}
\omega_4^{\nu_4}}{z_1^\ast},\label{eq:transf1}
\ee
corresponds to a set of distinct wave vectors in $\tilde{A}(\k)$,
\be
F_1(z_1,z_2) \leftrightarrow
\{\vec{c_1},\vec{c_2},\ldots,\vec{c_d}\}\subseteq\tilde{A}(\k)
\hspace{0.5in}d\leq p
\ee
determined by the distinct amplitudes in (\ref{eq:transf1}) i.e. $z_i$
corresponds to $\vec{k_i}$ and $z_i^\ast$ corresponds to $\vec{-k_i}$. For
example, $F_1(z_1,z_2)={\sigma_1}/{z_1^\ast}=z_1$
corresponds to $\{\vec{c_1}\}=\{\vec{k_1}\}$, and
$F_1(z_1,z_2)={\sigma_7}/{z_1^\ast}= (z_1^\ast)^{n-1}(z_2^\ast)^{n}(z_3z_4)^l$
corresponds to $\{\vec{c_1},\vec{c_2},\vec{c_3},\vec{c_4}\}=
\{\vec{-k_1},\vec{-k_2},\vec{k_3},\vec{k_4}\}$ if $n>1$ and
$\{\vec{c_1},\vec{c_2},\vec{c_3}\}= \{\vec{-k_2},\vec{k_3},\vec{k_4}\}$ if
$n=1$.
We construct
$P(\vec{k},\vec{k'_1},\vec{k'_2},\ldots,\vec{k'_d},\ldots,\vec{k'_p})$ by first
multiplying factors of $I_i$ with arguments that specify the inner products
amongst the arguments $(\vec{k},\vec{k'_1},\vec{k'_2},\ldots,\vec{k'_d})$:
\beq
\vec{k}\cdot\vec{k'_i}&=&\vec{k_1}\cdot\vec{c_i}\hspace{0.5in}i=1,\ldots d
\label{eq:prod1}\\
\vec{k'_i}\cdot\vec{k'_j}&=&\vec{c_i}\cdot\vec{c_j}\hspace{0.5in}i,j=1,\ldots
d.
\label{eq:prod2}
\eeq
The remaining arguments of $P$ (if required),
$\{\vec{k'_{d+1}},\ldots,\vec{k'_p}\}$, are specified to equal one of the
previous wave vectors $\{\vec{k'_{1}},\ldots,\vec{k'_d}\}$, e.g.
$I_1(\vec{k'_{1}},\vec{k'_{d+1}})$ sets $\vec{k'_{d+1}}=\vec{k'_{1}}$
and adds a factor of $a(\vec{k'_1})$.
In this way the monomial in (\ref{eq:transf1}) is built up. Note that in this
construction, $\{\vec{k'_{d+1}},\ldots,\vec{k'_p}\}$ are completely determined
by the arguments $\{\vec{k'_{1}},\ldots,\vec{k'_d}\}$. In particular,
when we set
$\{\vec{k'_{1}},\ldots,\vec{k'_d}\}=\{\vec{c_1},\vec{c_2},\ldots,\vec{c_d}\}$
this determines a set of wave vectors, $\{\vec{k'_{1}},\ldots,\vec{k'_p}\}=
\{\vec{c_1},\vec{c_2},\ldots,\vec{c_d},\vec{k'_{d+1}},\ldots,\vec{k'_p}\}$,
that
necessarily satisfy
\be
\vec{k}_1=\vec{c_1}+\vec{c_2}+\ldots+\vec{c_d}+\vec{k'_{d+1}}\ldots+\vec{k'_p}.
\label{eq:c1cd}
\ee
This follows from the assumption that $z_1^\ast\,F_1$ in (\ref{eq:transf1}) is
$\tor$-invariant.

\item We now prove that this construction of an $O(2)$ invariant $P$ always
yields a function that satisfies the requirements of translation symmetry
(\ref{eq:c1}) as well. We begin with a useful reduction: since
$|\vec{k}|=\k=|\vec{k}_1|$ in (\ref{eq:c1}) we can always rotate the arguments
of $P$ so that $\vec{k}=\vec{k}_1$ using the $O(2)$ invariance of $P$:
\be
P(\vec{k},\vec{k'_1},\vec{k'_2},\ldots,\vec{k'_p}) =
P(\vec{k}_1,\rot{\phi_k}\cdot\vec{k'_1},\rot{\phi_k}\cdot\vec{k'_2},\ldots,
\rot{\phi_k}\cdot\vec{k'_p})
\ee
where $\phi_k$ is the angle such that $\rot{\phi_k}\cdot\vec{k}=\vec{k_1}$.
Thus we need only verify (\ref{eq:c1}) for the specific case
$\vec{k}=\vec{k}_1$:
\be
P(\vec{k}_1,\vec{k'_1},\vec{k'_2},\ldots,\vec{k'_p})=0\;\;\mbox{\rm unless}\;\;
\vec{k}_1=\vec{k'_1}+\vec{k'_2}+\ldots+\vec{k'_p}.\label{eq:c1k1}
\ee
For most sets of arguments
$(\vec{k}_1,\vec{k'_1},\vec{k'_2},\ldots,\vec{k'_p})$, $P=0$, so we need only
examine those sets of vectors for which $P\neq0$ and verify that
$\vec{k}_1=\vec{k'_1}+\vec{k'_2}+\ldots+\vec{k'_p}$ holds in those cases.
By construction $P\neq0$ when we set
$\{\vec{k'_{1}},\ldots,\vec{k'_d}\}=\{\vec{c_1},\vec{c_2},\ldots,\vec{c_d}\}$
and in this case (\ref{eq:c1cd}) provides the desired relation. In general,
however, this is not the only set of wave vectors
$\{\vec{k'_{1}},\ldots,\vec{k'_d}\}$ in ${A}(\k)$ selected by the prescribed
inner products in (\ref{eq:prod1}) - (\ref{eq:prod2}), and it is necessary to
verify
(\ref{eq:c1k1}) for all the allowed sets $\{\vec{k'_{1}},\ldots,\vec{k'_d}\}$
for which $P\neq0$.

\item The occurrence of additional selected sets of vectors is due to the
reflection $\gamma_{k_1}\cdot\vec{k_1}=\vec{k_1}$ that fixes the vector
$\vec{k_{1}}$; in terms of $\g 3$ and the angle,
$\tan\theta_{ln}=n/l,$
we find
$\gamma_{k_1}=\rot{\theta_{ln}}\g 3\rot{-\theta_{ln}}=\rot{2\theta_{ln}}\g 3$
where $\g 3$ is the reflection defined in (\ref{eq:g3}).
Setting $\vec{k}=\vec{k_1}$ in (\ref{eq:prod1}), the equation for $\vec{k'_i}$
(for fixed $i$)
$\vec{k_1}\cdot\vec{k'_i}=\vec{k_1}\cdot\vec{c_i}$
has exactly two solutions in ${A}(\k)$:  $\vec{k'_i}=\vec{c_i}$ and
$\vec{k'_i}=\gamma_{k_1}\cdot\vec{c_i}$.
If $\vec{c_i}\neq\gamma_{k_1}\cdot\vec{c_i}$, then for $\vec{k}=\vec{k_1}$
there is a second set of vectors
selected by the inner products (\ref{eq:prod1}) - (\ref{eq:prod2}) for which
$P\neq0$:
\be
\{\vec{k'_{1}},\ldots,\vec{k'_d}\}=
\{\gamma_{k_1}\cdot\vec{c_1},\gamma_{k_1}\cdot\vec{c_2},\ldots,
\gamma_{k_1}\cdot\vec{c_d}\}.
\ee
For this second set, we obtain
\be
\vec{k}_1=\gamma_{k_1}\cdot\vec{c_1}+ \gamma_{k_1}\cdot\vec{c_2}+\ldots+
\gamma_{k_1}\cdot\vec{c_d}+\gamma_{k_1}\cdot\vec{k'_{d+1}}\ldots+
\gamma_{k_1}\cdot\vec{k'_p}
\label{eq:gc1cd}
\ee
by applying $\gamma_{k_1}$ to (\ref{eq:c1cd})
This completes the verification of (\ref{eq:c1k1}), and establishes the
translation symmetry of $P$.
{\bf $\Box$}
\end{enumerate}
\end{quote}
This lemma completes the proof of Theorem II.1.

\subsection{Discussion}

In summary, if $\fix\per={E_{\k^2}(\ext)}$ carries an irreducible
representation of the normalizer, then the restriction of Euclidean symmetric
maps on ${E_{\k^2}(\rtwo)}$ to the subspace ${E_{\k^2}(\ext)}$ yields the
entire module $\vec{M}(\semi)$. These cases are summarized in Table 4. Thus the
Euclidean symmetry of $\strobe$ does not imply constraints on
$\Ft=\strobe|_{\fix\per}$ beyond the requirement of $\semi$-equivariance. In
these cases, the hidden rotational symmetry has no consequences.

In the context of codimension-one bifurcations, where the assumption that
${E_{\k^2}(\rtwo)}$ carries an irreducible representation of $\euc$ is
appropriate, there are values of $\k^2$ for which $\fix\per$ fails to be
$\semi$-irreducible. As $\k^2$ increases, the first example of this occurs for
$\k^2=25$ where the modes $\vec{k}_1=(5,0)$, $\vec{k}_2=(0,5)$,
$\vec{k}_3=(4,3)$, $\vec{k}_4=(3,4)$ are degenerate due to the hidden rotations
and ${E_{25}(\rtwo)}=V_1\oplus V_2$ decomposes into two $\semi$-irreducible
subspaces with $V_1$ spanned by the modes $(\vec{k}_1,\vec{k}_2)$ and $V_2$
spanned by the modes $(\vec{k}_3,\vec{k}_4)$. This decomposition is discussed
at greater length elsewhere.\cite{cgl} The effect of the hidden rotational
symmetry on such a bifurcation will be studied in a separate paper.\cite{craw4}
In the context of codimension-two bifurcations, one can vary parameters to
locate mode interactions where the critical eigenspace
${E_{\k_1^2}(\rtwo)}\oplus{E_{\k_2^2}(\rtwo)}$ is characterized by two unequal
values of $\k^2$ and the representation of $\euc$ is reducible.  This will
necessarily lead to a reducible representation of $\semi$ on $\fix\per$ and
thus allow at least the possibility that the hidden rotations could have
observable consequences. Investigation of particular mode interactions is thus
another strategy of discovering bifurcations that can reveal the effects of
hidden rotations.\cite{craw4}
\begin{table}[h]
\begin{center}
Table 4.  Normalizer symmetry $\norm\per/\per$ and the module of
$\norm\per$-symmetric maps on $E_{\k^2}(\ext)$ obtained by the restriction
$\strobe |_{{E_{\k^2}(\ext)}}$
\end{center}
\vspace{7mm}
\begin{tabular}{lcccc}
\underline{$(l,n)$}
&\underline{$E_{\k^2}(\ext)$}
&\underline{$\norm\per/\per$ on $E_{\k^2}(\ext)$}
&\underline{Representation}
&\underline{$\{\strobe |_{{E_{\k^2}(\ext)}}\}$}\\
\\
$l=n>0$ & ${\bf R}^4$ & $\semi$ & irreducible& $\vec{M}(\semi)$\\
\\
$l>n=0$& ${\bf R}^4$ & $\semi$& irreducible & $\vec{M}(\semi)$\\
\\
$l>n>0$ & ${\bf R}^8$ & $\semi$& irreducible & $\vec{M}(\semi)$\\
\end{tabular}
\par
\vspace{7 mm}
\underline{Notes:}
\par $\{\strobe |_{{E_{\k^2}(\ext)}}\}$ represents the set of all maps on
$E_{\k^2}(\ext)$ obtained by the restriction (\ref{eq:pbcres}) from
$E_{\k^2}(\rtwo)$. When the normalizer $\norm\per$ acts irreducibly, the set of
$\semi$-symmetric maps obtained by restriction is a module of maximum size:
$\vec{M}(\norm\per)$.
\protect \vspace*{\fill}
\end{table}

\section{\hspace{0.125in}Neumann boundary conditions and hidden translation
symmetry}
Our discussion has shown that if the normalizer $\norm\per$ acts irreducibly on
eigenspace ${E_{\k^2}(\ext)}$, then the restriction of Euclidean symmetric maps
on ${E_{\k^2}(\rtwo)}$ to the subspace ${E_{\k^2}(\ext)}$ yields precisely the
module $\vec{M}(\semi)$. This is the module of symmetric maps associated with
the normalizer $\norm\per/\per$ and is the largest collection of maps that one
could hope to obtain by this restriction. For these same values of $\k^2$, we
now consider the further restriction of $\vec{M}(\semi)$ to the subspace
$\fix\neu={E_{\k^2}(\sg)}$ which yields those maps consistent with Neumann
boundary conditions on $\sg$.

\subsection{Restriction to $\fix\neu$}

This second restriction is quite analogous to the first, and one might expect
to obtain the module $\vec{M}(\norm\neu)$ of $\norm\neu$-symmetric vector
fields on ${E_{\k^2}(\sg)}$. This expectation is borne out for the
four-dimensional irreducible representations of $\semi$ but an exception occurs
for the  eight-dimensional representation when $(l+n)$ is even.
\begin{table}[h]
\begin{center}
Table 5.  Normalizer symmetry on $\fix\neu={E_{\k^2}(\sg)}$
\end{center}
\vspace{7mm}
\begin{tabular}{lccc}
\underline{$(l,n)$}&\underline{$E_{\k^2}(\ext)$}
&\underline{$\fix\neu\subset E_{\k^2}(\ext)$}
&\underline{$\norm\neu/\neu$ on $\fix\neu$}\\
\\
$l=n>0$ & $(z_1,z_2)\approx{\bf R}^4$ & $(u,u)\approx{\bf R}$ & $Z_2(-I)$\\
\\
$l>n=0$ & $(z_1,z_2)\approx{\bf R}^4$ & $(u_1,u_2)\approx{\bf R}^2$ & $\{\g 2,
\tran{\pm\pi/l}0 \}\approx\sgd$ \\
\\
$l>n>0$ & $(z_1,z_2,z_3,z_4)\approx{\bf R}^8$ & $(u_1,u_1,u_2,u_2)\approx{\bf
R}^2$ & $\left\{\begin{array}{cc}
\{\g 2,\tran {\pm\pi}0\}\approx\sgd& (l+n) \;\;\mbox{\rm odd} \\
\\
Z_2(\g 2)\times Z_2(-I)& (l+n)\;\;\mbox{\rm even}\end{array}\right.$ \\
\end{tabular}
\par
\vspace{7 mm}
\underline{Notes:}
\par
In the third column  $u,u_1$, and $u_2$ are real coordinates. In the fourth
column  $\{\g 2,\tran {\pm\pi/l}0\}$ denotes the subgroup of $\semi$ generated
by $\g 2$ and $\tran {\pm\pi/l}0$ whose action on $\fix\neu$ is isomorphic to
the two-dimensional representation of $\sgd$; similarly $\{\g 2,\tran
{\pm\pi}0\}$  is the subgroup generated by $\g 2$ and $\tran {\pm\pi}0$ which
is isomorphic to the same two-dimensional representation of $\sgd$.
\protect \vspace*{\fill}
\end{table}

In this exceptional case  the  normalizer $\norm\neu$ acts reducibly on
${E_{\k^2}(\sg)}$, and we only obtain a sub-module of $\vec{M}(\norm\neu)$. The
reducibility of $\norm\neu$ is clearly significant:  when ${E_{\k^2}(\ext)}$
carries an irreducible representation of $\norm\per$, then a reducible
representation of $\norm\neu$ necessarily implies that some maps in
$\vec{M}(\norm\neu)$ cannot be obtained by restriction from $\vec{M}(\semi)$.

In all cases considered ${E_{\k^2}(\sg)}$ is either one or two-dimensional,
c.f. Table 5, and  the quotient $\norm\neu/\neu$ depends on $(l,n)$. In the
one-dimensional case $l=n$, $\norm\neu/\neu$ is $Z_2(-I)$. In the
two-dimensional cases when $l>n=0$ or $l>n>0$ with $(l+n)$ odd, the
representation of $\norm\neu$ is irreducible and $\norm\neu/\neu$ is isomorphic
to the two-dimensional representation of $\sgd$. This representation is
generated on $(u_1,u_2)\in{\bf R}^2$  by the two reflections
\beq
(u_1,u_2)&\rightarrow&(u_2,u_1)\label{eq:g2gen}\\
(u_1,u_2)&\rightarrow&(u_1,-u_2);
\eeq
the first generator (\ref{eq:g2gen}) represents the diagonal reflection $\g 2$.
When $l>n>0$ with $(l+n)$ even, then $\norm\neu$ acts reducibly and
$\norm\neu/\neu$ is isomorphic to $Z_2(\g 2)\times Z_2(-I)$ where $Z_2(\g 2)$
is generated by (\ref{eq:g2gen}).

For the two-dimensional irreducible case the ring $R(\sgd)$ and module
$\vec{M}(\sgd)$ are well known.\cite{craw3,gss} The ring $R(\sgd)=\{N,\Delta\}$
is generated by the invariants,
$N=u_1^2+u_2^2$ and $\Delta=\delta^2$
where
$\delta=u_2^2-u_1^2,$
and $f(u_1,u_2)\in\vec{M}(\sgd)$ can always be expressed in the form
\be
f(u_1,u_2)=P(N,\Delta)\,\left(\begin{array}{c}u_1\\u_2\end{array}\right)
+Q(N,\Delta)\,\delta\left(\begin{array}{c}u_1\\-u_2\end{array}\right).
\label{eq:d44d}
\ee
Here  $P(x,y)$ and $Q(x,y)$ are arbitrary smooth real-valued functions.
For the reducible representation of $\norm\neu/\neu=Z_2(\g 2)\times Z_2(-I)$,
there is a second quadratic invariant
$\eta=u_1u_2,$
and an arbitrary $Z_2(\g 2)\times Z_2(-I)$-equivariant has the form
\be
f(u_1,u_2)= P(N,\eta)\,\left(\begin{array}{c}u_1\\u_2\end{array}\right)+
Q(N,\eta)\,\left(\begin{array}{c}u_2\\u_1\end{array}\right).\label{eq:nneu}
\ee

\begin{theorem} Assume $E_{\k^2}(\ext)=\fix\per$ carries an irreducible
representation of $\semi$, and let $\vec{M}(\semi)$ denote the module of
$\semi$-symmetric vector fields on $E_{\k^2}(\ext)$. Let
$\vec{M}(\semi)|_{\fix{\neu}}$ denote the set of maps obtained by restricting
this module to $\fix{\neu}$, and let $\vec{M}(\norm\neu)$ denote the module of
$\norm\neu$-symmetric vector fields on $\fix{\neu}$. Denote the two-dimensional
representation in {\rm (\ref{eq:g2gen})} by $\sgd$. Then
$\vec{M}(\semi)|_{\fix{\neu}}$ is a module of $\norm\neu$-symmetric vector
fields. The size of this module depends on the representation of $\semi$
carried by $E_{\k^2}(\ext)$.
\begin{enumerate}
\item Four-dimensional representations of $\semi$:
\begin{enumerate}
\item for $l=n>0$,  $\norm\neu=Z_2(-I)$ and
$\vec{M}(\semi)|_{\fix{\neu}}=\vec{M}(\norm\neu)$;
\item for $l>n=0$,  $\norm\neu=\sgd$ and
$\vec{M}(\semi)|_{\fix{\neu}}=\vec{M}(\norm\neu)$.
\end{enumerate}
\item Eight-dimensional representations of $\semi$:
\begin{enumerate}
\item for $l>n>0$ and $(l+n)$ odd,  $\norm\neu/\neu=\sgd$ and
$\vec{M}(\semi)|_{\fix{\neu}}=\vec{M}(\norm\neu)$;
\item for $l>n>0$ and $(l+n)$ even,  $\norm\neu/\neu=Z_2(\g 2)\times Z_2(-I)$
and $\vec{M}(\semi)|_{\fix{\neu}}$ is a submodule of $\vec{M}(\norm\neu)$.
This submodule is generated over the ring $R(\sgd)=\{N,\Delta\}$ by the four
$Z_2(\g 2)\times Z_2(-I)$ equivariants:
\be
\begin{array}{cc}
E_1(u_1,u_2)=\left(\begin{array}{c}u_1\\u_2\end{array}\right) &
E_2(u_1,u_2)=\delta\left(\begin{array}{c}u_1\\-u_2\end{array}\right)\\
\\
E_3(u_1,u_2)=\eta^l\left(\begin{array}{c}u_1\\u_2\end{array}\right) &
E_4(u_1,u_2)=\eta^{l-1}\left(\begin{array}{c}u_2\\u_1\end{array}\right).
\end{array}
\ee
That is, $\vec{M}(\semi)|_{\fix{\neu}}$ is the collection of vector fields of
the form
\beq
f(u_1,u_2)&=&
A_1(N,\Delta)\,E_1(u_1,u_2)+A_2(N,\Delta)\,E_2(u_1,u_2)\nonumber\\
&&\;\;\;\;
+A_3(N,\Delta)\,E_3(u_1,u_2)+A_4(N,\Delta)\,E_4(u_1,u_2)\label{eq:evmod}
\eeq
where the $A_i(x,y),\;\;i=1,2,3,4,$  are arbitrary smooth real-valued
functions.
\end{enumerate}
\end{enumerate}
 \end{theorem}
\noindent {\em {\bf Proof}.}
\begin{enumerate}
\item For each of the four-dimensional representations, Gomes has shown that
the vector fields in $\vec{M}(\semi)$ may be written as
\be
\Ft(z_1,z_2)= \tilde{p}(\tilde{N},\tilde{\rho}) \left(\begin{array}{c}
z_1\\z_2\end{array}\right) + \tilde{q}(\tilde{N},\tilde{\rho})
\left(\begin{array}{c} z_1|z_2|^2\\z_2|z_1|^2\end{array}\right) +
\tilde{r}(\tilde{N},\tilde{\rho}) \left(\begin{array}{c}
z_1|z_1|^2\\z_2|z_2|^2\end{array}\right)\label{eq:llnf2}
\ee
where
$\tilde{N}=|z_1|^2 + |z_2|^2$ and $\tilde{\rho}=|z_1|^2|z_2|^2 $
provide a Hilbert basis for the $\semi$ invariants and  $\tilde{p}$,
$\tilde{q}$, $\tilde{r}$ are arbitrary real-valued functions of these basic
$\semi$ invariants.\cite{gomes} For the case $l=n>0$, we set $(z_1,z_2)=(u,u)$
to restrict to the one-dimensional subspace $\fix\neu$, and $\Ft$ restricts to
$f(u)$:
\be
f(u)=\Ft|_{(z_1,z_2)=(u,u)}=p(u^2)\;u\label{eq:z2}
\ee
with $p(u^2)=\tilde{p}(2u^2,u^4) + \tilde{q}(2u^2,u^4) +\tilde{r}(2u^2,u^4)$.
Since any smooth function $p(x)$ can arise in this way, all maps of the form
(\ref{eq:z2}) are obtained. This is precisely the module $\vec{M}(Z_2(-I))$, of
reflection-symmetric maps in one dimension, or equivalently the module of maps
having $\norm\neu$ symmetry since $\norm\neu/\neu=Z_2(-I)$ for this
representation, cf. Table 5. For the case $l>n=0$, we set $(z_1,z_2)=(u_1,u_2)$
to restrict to the two-dimensional subspace $\fix\neu$, and $\Ft$ restricts to
$f(u_1,u_2)$:
\beq
f(u_1,u_2)&=&\Ft|_{(z_1,z_2)=(u_1,u_2)}\nonumber\\
&=&\tilde{p}({N},{\rho})\; \left(\begin{array}{c}
u_1\\u_2\end{array}\right) + \tilde{q}({N},{\rho})\;\left(\begin{array}{c} u_1
u_2^2\\u_2 u_1^2\end{array}\right) +
\tilde{r}({N},{\rho}) \left(\begin{array}{c}
u_1^3\\u_2^3\end{array}\right)
\eeq
where $N=u_1^2+u_2$ and $\rho=(N^2-\Delta)/4$. Using the relations
\beq
\left(\begin{array}{c} u_1 u_2^2\\u_2 u_1^2\end{array}\right) &=&
\frac{N}{2}\left(\begin{array}{c}u_1\\u_2\end{array}\right) +
\frac{3\delta}{2}\left(\begin{array}{c}u_1\\-u_2\end{array}\right)\\
&&\nonumber\\
\left(\begin{array}{c}u_1^3\\u_2^3\end{array}\right) &=&
\frac{N}{2}\left(\begin{array}{c}u_1\\u_2\end{array}\right) +
\frac{\delta}{2}\left(\begin{array}{c}u_1\\-u_2\end{array}\right)
\eeq
$f(u_1,u_2)$ can be re-expressed as (\ref{eq:d44d})
where $P(N,\Delta)=\tilde{p}({N},{\rho})+N[\tilde{q}({N},{\rho}) +
\tilde{r}({N},{\rho})]/2$ and $Q(N,\Delta)=[3\tilde{q}({N},{\rho}) +
\tilde{r}({N},{\rho})]/2$. Again, any smooth functions $P(x,y)$ and $Q(x,y)$
can arise in this construction so we obtain all maps of the form
(\ref{eq:d44d}). This is the module of $\sgd$-symmetric maps for the
two-dimensional irreducible representation of $\sgd$; since
$\norm\neu/\neu=\sgd$ for this case, the maps in (\ref{eq:d44d}) give precisely
the module associated with the normalizer. This proves the first part of the
theorem.

\item The proof of the second part of the theorem is more complicated, in part
because we lack an explicit set of generators for the module $\vec{M}(\semi)$
of $\semi$-symmetric vector fields for the eight-dimensional representation. We
first note that for $l>n>0$, the $\sgd$ invariants $(N,\Delta)$ and
equivariants $E_1(u_1,u_2)$ and $E_2(u_1,u_2)$ always extend to  $\semi$
invariants and equivariants on $V$. This can be checked by direct construction.
The $\semi$ invariants
$\tilde{N}\equiv\frac{1}{2}[|z_1|^2+|z_2|^2+|z_3|^2+|z_4|^2]$ and
$\tilde{\Delta}\equiv\frac{1}{4}[|z_3|^2+|z_4|^2-|z_1|^2-|z_2|^2]^2$
restrict to the $\sgd$ invariants on $\fix\neu$ when we set
$(z_1,z_2,z_3,z_4)=(u_1,u_1,u_2,u_2)$:
\be
(N,\Delta)=(\tilde{N},\tilde{\Delta})|_{\fix\neu},
\ee
and the $\semi$ equivariants
\be
\tilde{E}_1(\mbox{\bf z})\equiv\left(\begin{array}{c}z_1\\z_2\\z_3\\z_4
\end{array}\right)
\hspace{0.5in}
\tilde{E}_2(\mbox{\bf z})\equiv \tilde{N}
\left(\begin{array}{c}z_1\\z_2\\z_3\\z_4 \end{array}\right) -
2\left(
\begin{array}{c}z_1|z_1|^2\\z_2|z_2|^2\\z_3|z_3|^2\\z_4|z_4|^2
\end{array}\right)
\ee
similarly restrict  to the generators $E_i(u_1,u_2),\;\;i=1,2$,
\beq
E_1(u_1,u_2)&=&\tilde{E}_1|_{\fix\neu}=\left(\begin{array}{c}u_1\\u_2
\end{array}\right)\\
E_2(u_1,u_2)&=&\tilde{E}_2|_{\fix\neu}=\delta\left(\begin{array}{c}u_1\\-u_2
\end{array}\right).
\eeq
Thus all maps of the form
\be
f(u_1,u_2)=
A_1(N,\Delta)\,E_1(u_1,u_2)+A_2(N,\Delta)\,E_2(u_1,u_2)\label{eq:odd}
\ee
can be obtained regardless of the parity of $(l+n)$; this is precisely the
module of all $\sgd$ symmetric maps described in (\ref{eq:d44d}).  When $(l+n)$
is odd, then $\norm\neu/\neu=\sgd$ and the module in (\ref{eq:odd}) is
precisely $\vec{M}(\norm\neu)$; hence
\be
\vec{M}(\semi)|_{\fix{\neu}}=\vec{M}(\norm\neu)\;\;\;\mbox{\rm for}\;\;(l+n)
\;\;\;\mbox{\rm odd.}
\ee
However, when $(l+n)$ is even, then $\norm\neu/\neu= Z_2(\g 2)\times Z_2(-I)$
which is a subgroup of $\sgd$; in this case the module in (\ref{eq:odd}) is a
submodule of $\vec{M}(\norm\neu)$. It is also a submodule of the set of maps
$\vec{M}(\semi)|_{\fix{\neu}}$ obtained by restriction, that is we claim the
following inclusions:
$\vec{M}(\sgd)\subset\vec{M}(\semi)|_{\fix{\neu}}\subset\vec{M}(\norm\neu)
\;\;\;\mbox{\rm for}\;\;(l+n) \;\;\;\mbox{\rm even.}$
That $\vec{M}(\semi)|_{\fix{\neu}}$ is strictly larger than $\vec{M}(\sgd)$,
when $(l+n)$ is even, follows from the fact that the equivariants
$E_3(u_1,u_2)$ and $E_4(u_1,u_2)$ may be obtained by restriction in this case.
Recall the general form of $\semi$ equivariants (\ref{eq:8dmapll})
\be
\Ft({\bf z})=\left(\begin{array}{c} F_1({\bf z})\\ F_1(\g 3\cdot{\bf z})\\
F_1(\g 2\cdot{\bf z})\\ F_1({\gt 1\g 2\cdot\bf z})\end{array}\right)
\label{eq:8dmapm}
\ee
and let $z_1^\ast F_1({\bf z})=\sigma_6({\bf z})$ and $\sigma_1({\bf
z})\sigma_6({\bf z})$, using the form of $\sigma_6({\bf z})$ appropriate for
even parity, to  define two equivariants:
\beq
\tilde{E}_3(\mbox{\bf z})&\equiv& \left(\begin{array}{c}
\sigma_1({\bf z})\,\sigma_6({\bf z})/z_1^\ast \\
\sigma_1(\g 3\cdot{\bf z})\,\sigma_6(\g 3\cdot{\bf z})/z_2^\ast\\
\sigma_1(\g 2\cdot{\bf z})\,\sigma_6(\g 2\cdot{\bf z})/z_3^\ast\\
\sigma_1(\gt 1\g 2\cdot{\bf z})\,\sigma_6(\gt 1\g 2\cdot{\bf z})/ z_4^\ast
\end{array}\right)\nonumber\\
&&\nonumber\\
&=&
\left(\begin{array}{c}
z_1\,(z_1^\ast)^{(l+n)/2}z_2^{(l-n)/2}z_3^{(l+n)/2}z_4^{(l-n)/2}\\
\\
z_2\,z_1^{(l-n)/2}(z_2^\ast)^{(l+n)/2}z_3^{(l-n)/2}z_4^{(l+n)/2}\\
\\
z_3\,z_1^{(l+n)/2}(z_2^\ast)^{(l-n)/2}(z_3^\ast)^{(l+n)/2}(z_4^\ast)^{(l-n)/2}\\ \\
z_4\,(z_1^\ast)^{(l-n)/2}z_2^{(l+n)/2}(z_3^\ast)^{(l-n)/2}(z_4^\ast)^{(l-n)/2}
\end{array}\right)\\
&&\nonumber\\
\tilde{E}_4(\mbox{\bf z})&\equiv&\left(\begin{array}{c}
\sigma_6({\bf z})/z_1^\ast\\
\sigma_6(\g 3\cdot{\bf z})/z_2^\ast\\
\sigma_6(\g 2\cdot{\bf z})/z_3^\ast\\
\sigma_6(\gt 1\g 2\cdot{\bf z})/z_4^\ast
\end{array}\right)\nonumber\\
&&\nonumber\\
&=&
\left(\begin{array}{c}
(z_1^\ast)^{(l+n-2)/2}z_2^{(l-n)/2}z_3^{(l+n)/2}z_4^{(l-n)/2}\\
\\
z_1^{(l-n)/2}(z_2^\ast)^{(l+n-2)/2}z_3^{(l-n)/2}z_4^{(l+n)/2}\\
\\
z_1^{(l+n)/2}(z_2^\ast)^{(l-n)/2}(z_3^\ast)^{(l+n-2)/2}(z_4^\ast)^{(l-n)/2}\\
\\
(z_1^\ast)^{(l-n)/2}z_2^{(l+n)/2}(z_3^\ast)^{(l-n)/2}(z_4^\ast)^{(l-n-2)/2}
\end{array}\right).
\eeq
whose restriction to $\fix\neu$ yields $E_3$ and $E_4$:
\beq
E_3(u_1,u_2)&=&\tilde{E}_3|_{\fix\neu}=
\eta^l\left(\begin{array}{c}u_1\\u_2\end{array}\right)\\
E_4(u_1,u_2)&=&\tilde{E}_4|_{\fix\neu}=\eta^{l-1}\left(\begin{array}{c}u_2\\u_1\end{array}\right).
\eeq
Obviously $E_3$ and $E_4$ belong to $\vec{M}(\semi)|_{\fix{\neu}}$, however
since these equivariants are not symmetric with respect to
$(u_1,u_2)\rightarrow(u_1,-u_2)$ they lack $\sgd$ symmetry and are not in
$\vec{M}(\sgd)$.

\item These constructions show that the module in (\ref{eq:evmod}) is contained
in $\vec{M}(\semi)|_{\fix{\neu}}$; we now prove that every map in
$\vec{M}(\semi)|_{\fix{\neu}}$ can be written in the form (\ref{eq:evmod}). It
is sufficient to consider the form of a $\semi$-symmetric vector field
(\ref{eq:8dmapm}) when $z_1^\ast\,F_1({\bf z})$ is a general $\tor$-invariant
monomial (\ref{eq:second})
$z_1^\ast\,F_1({\bf z})=\sigma_1^{\nu'_1} \sigma_2^{\nu'_2} \sigma_3^{\nu'_3}
\sigma_4^{\nu'_4}\;
\omega_1^{\nu_1}\omega_2^{\nu_2}\omega_3^{\nu_3}\omega_4^{\nu_4}$
where we must have either $\nu_1'\geq1$ or $\nu_1\leq-1$ (or both).
The restriction of this vector field to $\fix\neu$ gives $f(u_1,u_2)$:
\be
f(u_1,u_2)=\Ft|_{(z_1,z_2,z_3,z_4)=(u_1,u_1,u_2,u_2)}=
\left(\begin{array}{c}
u_1^{q}\, u_2^{q'} \\ u_2^{q}\, u_1^{q'}\end{array}\right).\label{eq:reduce}
\ee
where
$q=2(\nu_1'+\nu_2')+|\nu_1|+|\nu_2|-1$ and
$q'=2(\nu_3'+\nu_4')+|\nu_3|+|\nu_4|.$
Inspection of the solutions (\ref{eq:nu1}) - (\ref{eq:nu2}) for
$(\nu_1,\nu_2,\nu_3,\nu_4)$ shows that
$|\nu_1|+|\nu_2|=\max\{|b|l,\,|a|n\}$ and
$|\nu_3|+|\nu_4|=\max\{|b|n,\,|a|l\};$
hence when $(a,b)$ are even, then $|\nu_1|+|\nu_2|$ and $|\nu_3|+|\nu_4|$ are
even implying that $f(u_1,u_2)$ has $\sgd$ symmetry and belongs to
$\vec{M}(\sgd)$. Thus we have only to examine the maps for $(a,b)$ odd, this
possibility only arises when $(l+n)$ is even and $(l,n)$ are each odd.  Thus we
can assume $|\nu_1|+|\nu_2|$ and $|\nu_3|+|\nu_4|$ are odd and therefore that
$q$ is even and $q'$ is odd. Since $q$ and $q'$ are unequal, either $q'>q$ or
$q>q'$. In the first instance, the map in (\ref{eq:reduce}) becomes
\be
\left(\begin{array}{c}
u_1^{q}\, u_2^{q'} \\ u_2^{q}\, u_1^{q'}\end{array}\right)
=\eta^{q-(l-1)}\;Z_1(J)
\ee
where $J=q'-q$ is odd and
\be
Z_1(J)=\eta^{l-1}\;\left(\begin{array}{c}
u_2^{J} \\ u_1^{J}\end{array}\right).
\ee
In the second instance ($q>q'$), the map in (\ref{eq:reduce}) becomes
\be
\left(\begin{array}{c}
u_1^{q}\, u_2^{q'} \\ u_2^{q}\, u_1^{q'}\end{array}\right)=\eta^{q'-l}\;Z_2(J')
\ee
where $J'=q-q'$ is odd and
\be
Z_2(J')=\eta^{l}\;\left(\begin{array}{c}
u_1^{J'} \\ u_2^{J'}\end{array}\right).
\ee
In each case the coefficients $\eta^{q-(l-1)}$ and $\eta^{q'-l}$, respectively,
are $\sgd$ invariant. Furthermore the equivariants $Z_1(J)$ and $Z_2(J')$ can
be expressed in terms of $E_3$ and $E_4$ with $\sgd$-invariant coefficients.
For low values of $(J,J')$ we have explicitly
$Z_1(1)=E_4(u_1,u_2)$, $Z_1(3)=N\,E_4(u_1,u_2)-E_3(u_1,u_2)$,
$Z_2(1)=E_3(u_1,u_2)$, and $Z_2(3)=N\,E_3(u_1,u_2)-\eta^2\,E_4(u_1,u_2)$;
the equivariants for $J\geq5$ and $J'\geq5$ can be reduced to these low order
cases using the identity
$Z(J)=N\,Z(J-2)-\eta^2\,Z(J-4)$
which holds for both $Z_1(J)$ and $Z_2(J')$. This establishes that each map in
$\vec{M}(\semi)|_{\fix{\neu}}$ is of the form (\ref{eq:evmod}).

\item Finally we check that the module (\ref{eq:evmod}) is indeed a submodule
of $\vec{M}(\norm\neu)$ when $(l+n)$ is even. Since $\norm\neu/\neu=Z_2(\g
2)\times Z_2(-I)$ in this case, generators for $\vec{M}(\norm\neu)$ are easily
found: any $f\in\vec{M}(\norm\neu)$ can be expressed in the form
(\ref{eq:nneu})
for arbitrary real-valued functions $P(x,y)$ and $Q(x,y)$. Here $Z_2(\g
2)\times Z_2(-I)$ symmetric terms lacking $\sgd$ symmetry can occur at any
order, however in (\ref{eq:evmod}) the lowest order terms lacking $\sgd$
symmetry are of degree $(2l-1)$ in $(u_1,u_2)$:
\be
\eta^{l-1}\left(\begin{array}{c}u_2\\u_1\end{array}\right).
\ee
This proves the second part of the theorem.

\end{enumerate}
\begin{quote}
{\bf $\Box$}\end{quote}
This theorem generalizes, and in one case corrects, an earlier analysis of the
restriction to $\fix\neu$ for these representations. In reference 8, some terms
in the restricted map (\ref{eq:evmod}) were overlooked.

\subsection{Discussion}

In the above theorem, as summarized in Table 6, two situations arise: either
\be
\vec{M}(\semi)|_{\fix\neu}=\vec{M}(\norm\neu)
\ee
or $\vec{M}(\semi)|_{\fix\neu}$ is strictly smaller than $\vec{M}(\norm\neu)$.
In the latter case, it is clear that the $\semi$ symmetry of $\Ft$ imposes
nontrivial constraints on $f=\Ft|_{\fix\neu}$ beyond the requirement of
$\norm\neu$-equivariance. These bifurcations are thus likely to show effects of
the hidden $\semi$ symmetry. Some of these effects have been discussed
elsewhere.\cite{craw3}
\begin{table}
\begin{center}
Table 6.  Normalizer symmetry $\norm\neu/\neu$ and the module of
$\norm\neu$-symmetric maps on $E_{\k^2}(\sg)$ obtained by the restriction
$\vec{M}(\semi) |_{{E_{\k^2}(\sg)}}$
\end{center}
\vspace{7mm}
\begin{tabular}{lcccc}
\underline{$(l,n)$}
&\underline{$E_{\k^2}(\sg)$}
&\underline{$\norm\neu/\neu$}
&\underline{Representation}
&\underline{$\vec{M}(\semi) |_{{E_{\k^2}(\sg)}}$}\\
\\
$l=n>0$ & ${\bf R}$ & $Z_2(-I)$ & irreducible & $\vec{M}(Z_2(-I))$\\
\\
$l>n=0$& ${\bf R}^2$ & $\sgd$ & irreducible &  $\vec{M}(\sgd)$\\
\\
$l>n>0$ & ${\bf R}^2$ & $\left\{\begin{array}{cc}
\sgd& (l+n) \;\;\mbox{\rm odd} \\
\\
Z_2^2& (l+n)\;\;\mbox{\rm even}\end{array}\right.$ & $\begin{array}{c}
\mbox{\rm irreducible}\\
\\
\mbox{\rm reducible }\end{array}$ & $\begin{array}{c} \vec{M}(\sgd)\\
\\
\subset\vec{M}(Z_2^2) \end{array}$
\\
\end{tabular}
\par
\vspace{7 mm}
\underline{Notes:}
\par
Here $Z_2^2$ denotes $Z_2(\g 2)\times Z_2(-I)$. When the normalizer $\norm\neu$
acts irreducibly, the module obtained by restriction has maximum size:
$\vec{M}(\norm\neu)$; when the normalizer acts reducibly we obtain only a
submodule of $\vec{M}(\norm\neu)$.
\protect \vspace*{\fill}
\end{table}

However, even if $\vec{M}(\norm\per|_{\fix\neu})=\vec{M}(\norm\neu)$, the
bifurcation of surface waves described by $f$ can reveal the hidden $\semi$
symmetry. This possibility arises because $\norm\neu$ acting on
${E_{\k^2}(\sg)}$ can be a larger group than the {\em geometric} symmetry group
of $\Omega$. Since $\Omega$ is a square domain its geometric symmetry group
$\Gamma_\Omega=\{\g 1,\g 2\}$ is generated by the two reflections
$\g 1\cdot(x,y)\rightarrow(\pi-x,y)$ and $\g 2\cdot(x,y)\rightarrow(y,x);$
this group is always contained in the normalizer
$\Gamma_\Omega\subseteq\norm\neu$. In the absence of hidden symmetry, one would
expect the bifurcation to be described by a map $f\in\vec{M}(\Gamma_\Omega)$
where $\vec{M}(\Gamma_\Omega)$ is the module of $\Gamma_\Omega$-symmetric
vector fields on ${E_{\k^2}(\sg)}$. As we have discussed elsewhere, for the
four-dimensional representations of $\semi$, $\Gamma_\Omega$ is a subgroup of
$\norm\neu$, and the module $\vec{M}(\norm\neu)$ is a submodule of
$\vec{M}(\Gamma_\Omega)$.\cite{craw3} Thus in these cases the effects of hidden
$\semi$ symmetry are also present. The recent experiments of Gollub and Lane
studied the bifurcations for $(l,n)=(2,0)$ and detected the predicted effects
of hidden $\semi$ symmetry. A detailed discussion of that work can be found in
reference 2.

\section{Acknowledgements}

I have enjoyed helpful conversations with M. Golubitsky and I. Melbourne; in
particular I am grateful to Melbourne for making available his unpublished
notes on the proofs of theorems I.1 and I.2. in the introduction. This work
was supported by the National Science Foundation under Grant DMS 9201028.


\newpage
\begin{thebibliography}{99}
\bibitem{sim} F. Simonelli and J.P. Gollub, Surface wave mode
interactions: effects of symmetry and degeneracy, {\em J. Fluid
Mech.} {\bf 199} (1989) 471.

\bibitem{cgl}  J.D. Crawford, J.P. Gollub, and David Lane, Hidden symmetries of
parametrically forced waves,  {\em Nonlinearity} {\bf 6} (1993) 119-164.

\bibitem{gms} M. Golubitsky, J.E. Marsden, and D.G. Schaeffer, Bifurcation
problems with hidden symmetries, in {\bf Partial Differential Equations and
Dynamical Systems} W.E. Fitzgibbon III, ed., Research Notes in Math. {\bf 101},
Pitman, San Francisco, (1984), p. 181.

\bibitem{fuj} H. Fujii, M. Mimura, and Y. Nishiura, A picture of
the global bifurcation diagram in ecological interacting and
diffusing systems, {\em Physica D} {\bf 5} (1982) 1.

\bibitem{arm1} D. Armbruster and G. Dangelmayr, Coupled
stationary bifurcations in non-flux boundary value problems, {\em
Math. Proc. Camb. Phil. Soc.} {\bf 101} (1987) 167.

\bibitem{arm2} G. Dangelmayr and D. Armbruster, Steady state mode
interactions in the presence of $O(2)$ symmetry and in non-flux
boundary conditions, in {\em Multiparameter Bifurcation Theory}
(eds. M. Golubitsky and J. Guckenheimer), Contemp. Math. {\bf 56}
(1986) 53.

\bibitem{craw1} J.D. Crawford, M. Golubitsky, M.G.M. Gomes, E.
Knobloch and I. Stewart, Boundary conditions as symmetry
constraints, in {\em Singularity Theory and Its Applications,
Warwick 1989}, vol. 2, (eds. R.M. Roberts and I.N.Stewart),
Lecture Notes in Mathematics, Springer-Verlag, Heidelberg (1991).

\bibitem{craw3} J.D. Crawford, Normal forms for driven surface waves:
boundary conditions, symmetry, and genericity, {\em Physica D} {\bf 52} (1991)
429-457.

\bibitem{mel} I. Melbourne, private communication.

\bibitem{gss} M. Golubitsky, I. Stewart, and D.G. Schaeffer, {\bf
Singularities and Groups in Bifurcation Theory}: Vol. II, Appl.
Math. Sci. {\bf 69}, Springer-Verlag, New York (1988).

\bibitem{dg} B. Dionne and M. Golubitsky, Planforms in two and three
dimensions, {\em ZAMP} {\bf 43} (1992) 36.

\bibitem{gomes} M.G.M. Gomes, Steady-state mode interactions in
rectangular domains, M.Sc. Thesis, Mathematics Institute, University of
Warwick, 1989.

\bibitem{gomes2} M.G.M. Gomes, Symmetries in bifurcation theory: the
appropriate context, Ph.D. Thesis, Mathematics Institute, University of
Warwick, 1992.

\bibitem{craw4} J.D. Crawford, ${D_4}\dot{+} T^2$ mode interactions and hidden
rotational symmetry, (1993) preprint.

\end{thebibliography}
\end{document}